# The Chemical Evolution of the Solar Neighbourhood: the Effect of Binaries.

E. De Donder [1] and D. Vanbeveren [2]

Astrophysical Institute, Vrije Universiteit Brussel, Pleinlaan 2, 1050 Brussel

*Abstract.* In this paper we compute the time evolution of the elements ($^{4}$He, $^{12}$C, $^{14}$N, $^{16}$O, $^{20}$Ne, $^{24}$Mg, $^{28}$Si, $^{32}$S, $^{40}$Ca and $^{56}$Fe) and of the supernova rates in the solar neighbourhood by means of a galactic chemical evolutionary code that includes in detail the evolution of both single and binary stars. Special attention is payed to the formation of black holes.

Our main conclusions:

- in order to predict the galactic time evolution of the different types of supernovae, it is essential to compute in detail the evolution of the binary population,
- the observed time evolution of carbon is better reproduced by a galactic model where the effect is included of a significant fraction of intermediate mass binaries,
- massive binary mass exchange provides a possible solution for the production of primary nitrogen during the very early phases of galactic evolution,
- chemical evolutionary models with binaries or without binaries but with a detailed treatment of the SN Ia progenitors predict very similar age-metallicity relations and very similar G-dwarf distributions whereas the evolution of the yields as function of time of the elements $^{4}$He, $^{16}$O, $^{20}$Ne, $^{24}$Mg, $^{28}$Si, $^{32}$S and $^{40}$Ca differ by no more than a factor of two or three,
- the observed time evolution of oxygen is best reproduced when most of the oxygen produced during core helium burning in ALL massive stars serves to enrich the interstellar medium. This can be used as indirect evidence that (massive) black hole formation in single stars and binary components is always preceded by a supernova explosion.

[1] ededonde@vub.ac.be
[2] dvbevere@vub.ac.be

# 1. Introduction.

Current chemical evolutionary models (CEMs) predict the observations in the solar neighbourhood within a factor of two (e.g. Timmes et al., 1995; Yoshii et al., 1996; Chiappini et al., 1997; Portinari et al., 1998; Samland, 1998; Boissier and Prantzos, 2000; Prantzos and Boissier, 2000). This putative success is astonishing since neither of these models includes in detail the effects of binary evolution. Even more astonishing is the fact that in most (all) of the studies where binaries are ignored, a (even qualitative) discussion on the effects of binaries is lacking. It should be clear that, since a significant fraction of the observed stars in the Galaxy are part of a binary system, a detailed study on the effects of binaries is necessary to make galactic chemical evolutionary calculations more reliable.

It is generally accepted that Type Ia supernovae (SN Ia) progenitors are exploding carbon-oxygen white dwarfs (CO WDs) in close binary systems. They are especially needed in CEMs to reproduce the observed time evolution of the abundance ratios of the α elements to iron. Therefore, most of the CEMs account for SN Ia's however, the fraction of close binary systems that produces them is in most cases presented as a free parameter (see Greggio & Renzini, 1983) that is fixed to meet some observational constraints, without modelling in detail the evolution of the progenitor systems and denying existing physically realistic binary evolutionary studies related to SN Ia events by invoking evolutionary uncertainties.

Many CEMs attempt to follow the evolution of the different types of SN although binaries are not included. However, the SN rates depend critically on the binary population (see also De Donder and Vanbeveren, 1998) and therefore the only realistic way to study the time evolution of the SN types in a galaxy is to combine a single star – binary population number synthesis (PNS) code with a SN model of any type and with a CEM describing the star formation history in a galaxy. In the first part of the paper (section 2), we summarise the overall PNS code used in Brussels. De Donder and Vanbeveren (1998) used this PNS code and discussed the SN II and SN Ib,c rates. Here we study in detail the expected SN Ia rates and the properties of the SN Ia binary progenitor population at birth. We do this by linking PNS and CEM with the two possible SN Ia models that have been proposed in literature: the single-degenerate scenario (Whelan & Iben, 1973; Nomoto, 1982; Hachisu et al., 1996, Li and van den Heuvel, 1997) and the double degenerate scenario (Iben & Tutukov, 1984; Webbink, 1984).

In the second part of the paper (section 3) we investigate how chemical evolutionary results are affected when the evolution of intermediate mass and massive binaries are taken into account in detail. The processes of Roche Lobe Overflow (RLOF) and mass accretion affect in a critical way the evolution of a star. In the case of massive stars, large mass loss or accretion of matter significantly modifies the formation of the carbon-oxygen (CO) core and consequently the ejection of heavy elements into the interstellar medium (ISM) during a supernova explosion. Changes in the chemical composition of surface layers caused by accretion of nuclear processed material may modify significantly the stellar wind (SW) yields. Intermediate mass binary components that accrete mass lost by a companion may turn into massive ones to produce more metals. Moreover, the post-core-helium burning evolutionary pattern which is typical for intermediate mass single stars and where newly formed carbon is mixed into the star's outer layers (carbon dredge up process) is not expected to happen when this star is the primary of an interacting close binary. This may obviously affect the overall carbon enrichment of the ISM.

Beside the influence of binary evolution we also look at the effects of black hole (BH) formation on chemical evolutionary calculations. As was already addressed by Maeder (1992), direct BH formation whereby no matter is ejected during the collapse, significantly reduces the metallicity production. Whether a massive star collapses directly into a BH or first has a SN outburst is at present unclear. Recent core collapse simulations indicate that BHs may form with a foregoing SN in the mass range $(25-40)M_\odot$ (Fryer, 1999). Also observational there is some evidence of BH formation with matter ejection (Israelian et al., 1999). We will demonstrate that the early phases of galactic chemical evolution depend critically on the physics of BH formation.

In a separate paper (De Donder and Vanbeveren, 2001) we will discuss the effects of binary parameters on chemical yields and we will outline in detail how our binary yields can be implemented in a CEM.

## 2. Population number synthesis (PNS).

### 2.1. General

A PNS model calculates the evolution as function of time of a stellar population consisting of single stars and interacting binaries of all types, with any mass ratio and orbital period, for starburst regions and for regions where star formation is continuous in time. Depending on the degree of sophistication of the PNS code it is possible to compute the evolution of the following single star and/or binary subpopulations: the main sequence stars, the giants, the supergiants, the Wolf-Rayet (WR) stars, the subdwarfs and the white dwarfs (WD's), the binaries with a compact companion [either a WD, a neutron star (NS) or a BH], the X-ray binaries (standard high mass and low mass X-ray binaries, soft sources and supersoft sources), the double compact star binaries. Most of the existing PNS codes are also able to estimate the population of binaries where both components merge. For the latter one obviously has to discriminate between binaries where both components are normal stars and binaries where at least one of the components is a compact star.

A detailed description of our PNS code can be found in Vanbeveren et al. (1998a, b, hereafter VB98) and Vanbeveren (2000, 2001). Our code was made in the first place to study massive star populations. However, it was straightforward to adapt the code so that it can handle intermediate mass single stars and binaries as well. In its present state it allows to explore a population of binaries with initial parameters $3 \leq M_1/M_\odot \leq 120$ ($M_1$ = primary mass), $0 < q \leq 1$ and 1 day $\leq P \leq$ 10 years, and a population of single stars with $0.1 \leq M/M_\odot \leq 120$. We use a detailed set of stellar evolutionary calculations for different initial metallicities ($0.001 \leq Z \leq 0.02$). In the case of binaries, it accounts for the Roche lobe overflow (RLOF) and mass transfer in case A and case $B_r$ binaries, the common envelope (CE) process and spiral-in in case $B_c$ and case C binaries, the CE process and spiral-in in binaries with a compact companion (a WD, NS or BH), the effect of an asymmetric SN explosion when a compact remnant is left, and, in the case of massive stars, the effect of SW mass loss where we use the most recent SW mass loss formalisms.

PNS results are sensitive to the process of convective core overshooting during stellar evolution. We use the formalism proposed by the Geneva group (Schaller et al., 1992). Remark that this differs significantly from the formalism preferred by the Padua group (Portinari et al., 1998).

The initial mass function $\varphi(M)$, the binary period distribution $\Pi(P)$ and the binary mass ratio distribution $\phi(q)$ are assumed to be constant in space and time.

For the solar neighbourhood $\varphi(M)$ satisfies a power-law with the following prescription:

$$\varphi(M) = \begin{cases} \varphi_1 \cdot M^{-2.3} & M \leq 2\, M_\odot \text{ (Salpeter, 1955)} \\ \varphi_2 \cdot M^{-2.7} & M > 2\, M_\odot \text{ (Scalo, 1986)} \end{cases} \quad (1)$$

with the appropriate normalisation. The mass fraction $\zeta$ of stars with $M \geq 2\, M_\odot$ is taken 0.3 (Pagel, 1980). The same $\varphi(M)$ holds for single stars and for primaries (the initially most massive component) of binaries.

$\Pi(P)$ is flat in log P (Popova et al., 1982; Abt, 1983) with P ranging from 1 day to 10 years and PNS (and CEM) calculations are performed for various $\phi(q)$ ($q = M_2/M_1$ = secondary mass/primary mass). We take either a flat distribution, a Hogeveen distribution that favours small q values (Hogeveen, 1991) or a $\phi(q) \sim q^{0.5}$ (Garmany, 1980) that favours binaries with $q \sim 1$.

When a SN explosion happens in a binary system one has to account for possible asymmetry by adding a kick velocity to the compact remnant star, a NS or a BH. The kick angles are isotropically distributed and the kick velocity distribution is taken similar to the 3D space velocity distribution of Galactic single radio pulsars that has been derived by Lorimer et al. (1997). These velocities imply an average kick velocity $\langle v_{kick} \rangle = 450$ km/s. To study the consequence of possible observational errors, we also perform test computations for a distribution with the same shape but adapted so that $\langle v_{kick} \rangle = 150$ km/s.

Below we list the evolutionary parameters that critically affect PNS predictions in general, SN rates in particular.

- Min(BH) = the initial mass for a massive star to collapse directly into a BH without a SN explosion. This mass limit is rather uncertain. For single stars we make our calculations with Min(BH) = 25 $M_\odot$ (= BH1) and with Min(BH) = 40 $M_\odot$ (= BH2). Since interacting binary components with initial mass $\leq$ 40 $M_\odot$ form CO cores with lower mass than their single star counterparts, we always consider case BH2 for the former. Stars with a zero-age mass smaller than Min(BH) may form low(er) mass BHs preceded by a SN outburst as indicated by the existence of runaway low mass X ray BHs (Nelemans et al., 1999). To know their mass we relate the final CO core masses obtained from our evolutionary computations with the computations of Woosley & Weaver (1995) (hereinafter WW95). The physics of explosive nucleosynthesis and the physics of the SN mechanism itself contain uncertainties which may affect galactic chemical evolutionary predictions. We will use the case B massive star SN models of WW95 but we will discuss the consequences of the latter uncertainties there where it is important.

- $\beta$ = the fraction of matter lost by the primary star during RLOF that is accreted by the secondary star. It is still a highly debated parameter and therefore, at present, we have to consider various possibilities. A probable formalism has been discussed in many studies on PNS (see also VB98) and we also use it here. Summarising:

  - components of binaries with an initial mass ratio q $\leq$ 0.2 evolve through a CE phase during RLOF and the low mass star will spiral-in in the envelope of the more massive companion. Most of these systems merge, some survive. Hydrodynamic simulations indicate that during the spiral in process accretion of mass onto the low mass star is very unlikely, i.e. $\beta$ = 0,
  - in case A/$B_r$ systems with q $\geq$ 0.4, the RLOF is accompanied by mass transfer but it is unclear at present whether or not all mass lost by the loser can be accreted by the gainer. We explore the consequences when 50% is transferred (and thus 50% is lost from the binary) or 100% ($\beta_{max}$ = 0.5 or 1). When matter leaves the system, the easiest way to do so is via the second

Lagrangian point. This allows us to calculate from first principles the orbital angular momentum loss and to determine the binary period variation,

- case $B_c$ and case C binaries evolve through a CE phase and no matter is accreted by the secondary, i.e. $\beta = 0$.

- $\alpha$ = the efficiency of the conversion of orbital energy into potential energy during CE and spiral-in. This efficiency may be different for non evolved binaries with an initial mass ratio $q \leq 0.2$ ($=\alpha_1$), case $B_c$/C binaries ($=\alpha_2$) and for spiral-in of a compact component (WD, NS or BH) ($=\alpha_3$).

Without any explicit specification, our calculations are performed for the following standard set of PNS parameters: BH2 for both single and interacting binary components, the WW95 case B massive star SN model, $\beta_{max} = 1$, $\alpha_1 = \alpha_2 = \alpha_3 = 1$, $<v_{kick}> = 450$ km/s and a Hogeveen initial mass ratio distribution.

## 2.2. The SN Ia model.

The SN II and SN Ib rates have been discussed by De Donder and Vanbeveren (1998). In the present section, we will focus on the SN Ia's.

PNS follows in detail the evolution of all binary systems. Therefore, once we agree upon a model for a Type Ia SN, PNS can calculate in a straightforward way the realisation frequencies. We like to remark here that this is the only physically realistic way to calculate these frequencies.

We consider separately the two most favoured binary scenarios i.e. the single degenerate (SD) scenario and the double degenerate (DD) scenario.

- SD scenario

Starting from a stellar generation on the ZAMS with a realistic binary population, our PNS code predicts in detail the content and binary properties of the population of WD +

companion star binaries. To decide whether a SN Ia happens or not, we link this predicted population with the SD results of Hachisu et al. (1999a). In this model the WD mass grows due to accretion of mass lost by its Roche lobe filling companion (we separately consider the WD binaries with a red giant = RG companion from those with a pre-RG companion) but the accretion process is regulated by an accompanying SW mass loss process: the higher the accretion the higher the stellar wind mass loss. Similarly as a massive star SW, the WD wind could be radiatively driven and in this case the WD wind could depend on the metallicity Z (Kobayashi et al., 1998). Kobayashi et al. (2000) try to find support for their metallicity dependent SD model by comparing predicted and observed SN Ia rates in different galaxies with a different Z. However, it is easy to demonstrate that *the predictions of a PNS where the SD is Z-dependent and the binary frequency is Z-independent are similar to those where the SD is Z-independent but the binary frequency is Z-dependent*. Therefore, when we use the SD model to estimate the chemical evolution (section 3), we will present calculations with the Z-dependent SD scenario but also with a Z-independent SD scenario. In the latter case, we use the SD results holding for Z = 0.02 for all Z.

Remark that the age of the SNIa's in the SD scenario is approximately the lifetime of the companion star at the beginning of mass transfer to the WD.

- DD scenario

Our PNS model predicts in detail the content and binary properties of the WD + WD binary population. The further evolution of a WD + WD binary is governed by orbital decay caused by gravitational wave radiation (GWR) and both WD's will merge after a characteristic timescale $\tau_{GWR}$ which is given by

$$\tau_{GWR} = 8 \cdot 10^7 \, (\text{yrs}) \cdot \left( \frac{(M_1 + M_2)^{\frac{1}{3}}}{M_1 \cdot M_2} \right) \cdot P^{\frac{8}{3}} \, (\text{hr}) \tag{2}$$

(Landau & Lifshitz, 1959).

In the DD scenario a SNIa explosion happens as a result of the merging of the two CO WD's provided that their combined mass is larger than the Chandrasekhar mass. It is obvious then to calculate the SN Ia rate as function of time. It should be clear that the age of the SNIa equals the evolutionary time of the system prior to the formation of both WDs plus the time of orbital decay under GWR.

Although the DD scenario offers a more easy way to approach the Chandrasekhar mass than the SD scenario, it is less preferred at present than the SD scenario. A main reason for this is that computations on WD mergers seem to indicate that an off-center ignition will convert carbon and oxygen into oxygen, neon, and magnesium leading to a gravitational collapse into a NS rather than to a thermonuclear disruption (e.g. Segretain et al., 1997; King et al., 2001).

**2.3. The SN Ia progenitor population**

To illustrate the differences between both scenario's, let us consider one generation of intermediate mass binary progenitors ($3 \leq M_{1,0}/M_\odot \leq 10$) (satisfying the appropriate distribution functions discussed earlier and taking Z=0.02). Our PNS code then computes the resulting SNIa properties when the SD and DD scenario apply. Table 1 gives the fraction A of all intermediate mass binaries that produces Type Ia SNe for different values of the PNS parameters. Note that in all cases, this fraction is very small (of the order of a few percent). Figure 1a illustrates the (normalised) age-frequency relation of the SN Ia resulting from one intermediate mass binary generation for both scenarios whereas figures 1b, 1c and 1d show the system characteristics of the progenitors on the zero-age main sequence (ZAMS). We like to remind that in interacting binaries, primaries with initial ZAMS mass up to ~10 $M_\odot$ develop degenerate CO-cores (for single stars this maximum mass ~ 8 $M_\odot$). This is illustrated in figure 1b.

The SN Ia formation depend, except for $\beta_{max}$, primarily on the adopted mass ratio distribution and on the efficiency factor $\alpha$ during CE evolution. Because SNIa's in the SD scenario come from rather low q-systems it is obvious that the overall production rate is minimal for a Garmany q-distribution that favours systems with a high initial mass ratio. Lowering the efficiency during CE reduces this rate since most progenitor systems evolve through CE phases whereby the orbital shrinkage can easily lead to a merger before one or both stars become degenerate. In our SD scenario the majority of the SNIa's arise from WD+pre-RG systems where the progenitor has undergone a CE phase. Most of the SNIa's happen within 0.3-1.3 Gyr after the formation of the progenitor binary generation. With the DD scenario SNIa's happen between 0.1 and 15 Gyr after the formation of the progenitor population with a maximum around 4 Gyr.

| Model | $\alpha_2=\alpha_3$ | $\beta_{max}$ | $\phi(q)$ | A (%) |
|---|---|---|---|---|
| SD | 1 | 1 | flat | 1.02 |
| SD | 1 | 1 | Hogeveen | 1.94 |
| SD | 1 | 1 | Garmany | 0.80 |
| SD | 0.5 | 1 | flat | 0.69 |
| SD | 1 | 0.5 | flat | 1.02 |
| DD | 1 | 1 | flat | 3.61 |
| DD | 1 | 1 | Hogeveen | 2.46 |
| DD | 1 | 1 | Garmany | 3.86 |
| DD | 0.5 | 1 | flat | 3.30 |
| DD | 1 | 0.5 | flat | 1.45 |

Table 1: The fraction A of all intermediate mass binaries that finally evolve into a Type Ia SN and this for the two considered scenarios combined with different mass ratio distributions, CE efficiency parameter and accretion parameter.

**Figure 1a.:** The age-frequency distribution of the SNIa's that result from one intermediate mass binary generation, as predicted by the SD model (black lines) and the DD model (grey lines) for the different types of mass ratio distribution i.e. a flat one (long dashed line), a Hogeveen one (full line) and a Garmany type (short dashed line), all combined with $\alpha_2=\alpha_3=1$ and $\beta_{max}=1$. The numbers on the vertical axis are the normalised realisation frequencies.

**Figure 1b**: The same as figure 1a but showing the distribution of the primary mass ($M_{1,0}$, in solar masses) of the progenitor systems on the ZAMS.

**Figure 1c**: The same as figure 1a but showing the distribution of the mass ratio ($q_0$) of the progenitor systems on the ZAMS.

**Figure 1d**: The same as figure 1a but showing the distribution of the logarithm of the orbital period ($P_0$, in days) of the progenitor systems on the ZAMS.

## 2.4. A special class of type II supernovae?

If the accretion rate is too low or too high for the WD to perform stable burning on its surface and to grow in mass, the WD undergoes either nova outbursts or starts to spiral-in into the envelope of the secondary. In the latter case, the system may merge. What happens during merging and what kind of object is finally formed is not known. It was suggested by Sparks & Stecher (1974) that if the combined mass of the WD and of the secondary's core exceeds the Chandrasekhar mass, the merging would possibly result in the formation of a NS whereby enough energy is released to power a SN explosion. We estimated their number with our PNS code and it is interesting to mention that if they indeed produce SNe (probably of Type II), they can make up a significant fraction (up to ~15%) of the total SNII population. In figures 2a, 2b and 2c we give the system characteristics of their progenitors on the ZAMS, when the standard set of PNS parameters applies. They are mainly short period systems with an intermediate mass ratio that undergo a partly conservative first RLOF. So, we want to report here that if WD with pre-RG/RG star mergers produce SNe then

- *a non-negligible fraction of Type II SNe may be related to intermediate mass close binaries and could be important for studies on the nature of SNe as well as for chemical evolutionary studies.*

**Figures 2a:** The distribution of the primary mass ($M_{1,0}$, in solar masses) on the ZAMS of the progenitor systems of merging WD + pre-RG/RG stars.

**Figures 2b:** The distribution of the mass ratio ($q_0$) on the ZAMS of the progenitor systems of merging WD + pre-RG/RG stars.

**Figures 2c:** The distribution of the logarithm of orbital period ($P_0$) on the ZAMS of the progenitor systems of merging WD + pre-RG/RG stars.

## 3. Chemical evolutionary computations with binaries.

### 3.1. The chemical evolutionary model.

In the present paper we will concentrate on the formation and evolution of the solar neighbourhood.

Similarly as was done in Chiappini et al. (1997), we assume that the Galaxy formation occurs in two phases of major infall, during which the halo-thick disk and the thin disk are formed respectively. The second phase starts roughly at the end of the first one. For the halo-thick disk and thin disk evolution, different accretion rates are adopted. The infalling gas is assumed to be of primordial chemical composition and occurs at the following rate:

$$\frac{dG_i(r,t)_{inf}}{dt} = \frac{A(r)}{\sigma(r,t_{now})} \cdot (X_i)_{inf} \cdot e^{-t/\tau_I} + \frac{B(r)}{\sigma(r,t_{now})} \cdot (X_i)_{inf} \cdot e^{-(t-t_{max})/\tau_{II}} \qquad (3)$$

with $G(r,t)_{inf}$ the normalised surface gas density of the infalling gas, $\sigma(r,t)$ the surface mass density, $t_{now}$ the present age of the Galaxy (~ 15 Gyr), $t_I$ and $t_{II}$ the time-scale for mass accretion in the halo-thick disk and thin disk respectively and $t_{max}$ the maximum period of gas accretion onto the disk which is taken to be 2 Gyr, marking the end of the halo-thick disk phase. The index i refers to the identity of the considered chemical element. For the solar neighbourhood (r = $r_\odot$ ≈ 10 kpc) $t_{II}$ = 8 Gyr and $t_I$ = 1 Gyr. The quantities A(r) and B(r) are fixed by the constraint of reproducing the current total surface mass densities of the thick and thin disk in the solar neighbourhood :

$$\begin{cases} A = \dfrac{\sigma_I(t_{now})}{\tau_I(1-e^{-t_{now}/\tau_I})} \\ B = \dfrac{\sigma(t_{now}) - \sigma_I(t_{now})}{\tau_{II}(1-e^{-(t_{now}-t_{max})/\tau_{II}})} \end{cases} \qquad (4)$$

with $\sigma(t_{now})$ and $\sigma_I(t_{now})$ respectively the present total surface mass density of the total and thick disk. We take $\sigma(t_{now}) = 50$ M$_\odot$ pc$^{-2}$ and $\sigma_I(t_{now}) = 10$ M$_\odot$ pc$^{-2}$ (Rana, 1991).

For simplicity often a Schmidt's star formation function (Schmidt, 1959, 1963) that only uses the surface density of gas, is used. However, the star formation process may be controlled by several processes complicating the star formation rate as function of place and time. Based on energy and dynamic considerations during the formation of the Galaxy, we use the following functional form (Talbot & Arnett, 1975; Chiosi, 1980):

$$\psi(r,t) = \tilde{v} \cdot \left[\frac{\sigma(r,t)}{\tilde{\sigma}(\tilde{r},t)}\right]^{2(k-1)} \cdot \left[\frac{\sigma(r,t_{now})}{\tilde{\sigma}(\tilde{r},t)}\right]^{(k-1)} \cdot G^k(r,t) \qquad (5)$$

with $\tilde{v}$ a parameter for the star formation efficiency and $\tilde{\sigma}(\tilde{r},t)$ the total surface mass density at a given galactocentric distance $\tilde{r}$ (here = 10 kpc) used as normalisation factor. If the surface gas density is below 7 M$_\odot$/pc$^2$ a star formation stop is assumed (Gratton et al., 1996).

The evolution of the chemical elements (He, C, N, O, Ne, Mg, S, Si, Ca, Fe) is followed in time in the solar neighbourhood by solving the following set of equations:

$$\frac{dG_i(t)}{dt} =$$
$$-X_i(t) \cdot \psi(t)$$
$$+ \int_{M_{min}}^{M_{max}} (1 - f_b(t-t_M)) \cdot \psi(t-t_M) \cdot \varphi(M) \cdot R_{M,i}(t-t_M) \cdot dM \qquad (6)$$
$$+ \int_{M_{min}}^{M_{max}} \int_{q_{min}}^{q_{max}} \int_{P_{min}}^{P_{max}} f_b(t-t_{M_1}) \cdot \varphi(M_1) \cdot \phi(q) \cdot \Pi(P) \cdot \left[\psi(t-t_{M_1}) \cdot R_{M_1,i}(t-t_{M_1}) + \psi(t-t_{M_2}) \cdot R_{M_2,i}(t-t_{M_2})\right] \cdot dP \cdot dq \cdot dM_1$$
$$+ \frac{dG_i(t)_{inf}}{dt}$$

t is the galactic timescale, $t_M$ is the evolutionary timescale of the star with mass M ($0 < t_M \leq$ total lifetime of the star), $R_{Mi}(t-t_M)$ is the mass fraction of a star of mass M formed at the moment ($t-t_M$) that is ejected at galactic time t (= stellar evolutionary time $t_M$) into the ISM in the form of element i. This ejection happens during the evolution of a star as a consequence of SW mass loss or as a consequence of the RLOF process when it is a binary component, and, at the end when a SN occurs. For single stars (second term in the equation), $R_{M,i}$ is a function of its initial ZAMS mass. In the case of a binary system (third term in the equation) we separate the mass fraction of the primary ($R_{M1,i}$) and the secondary ($R_{M2,i}$) star. The factor $f_b(t)$ is the binary formation rate as function of time. The integration limits are $[M_{min}, M_{max}]$=[0.1, 120]$M_\odot$, $[q_{min}, q_{max}]$=[0, 1], $[P_{min}, P_{max}]$=[1, 3650]days.

## 3.2. The binary formation rate $f_b$

The parameter $f_b$ stands for the formation rate of binaries with the properties given above (= the fraction of binaries with the properties given above on the zero age main sequence). It is easy to demonstrate that the majority of these binaries will interact, i.e. the primaries in most of these binaries will fill their Roche volume at a certain moment during their evolution. It cannot be excluded that $f_b$ is a function of mass, mass ratio and period but in all our models where the effect of binaries is included in detail we assume that $f_b$ is independent from the latter parameters.

From observational studies on spectroscopic binaries in the solar neighbourhood we know that about 33% (±13%) of the O stars are the primary of a massive close binary with a mass ratio q >0.2 and a period P ≤ 100 days (Garmany et al., 1980). A similar conclusion holds for the intermediate mass B type stars (Vanbeveren et al., 1998). Accounting for observational selection, it can be shown by binary population synthesis studies that to meet the above observations, an initial OB type binary fraction $f_b$ (= the binary formation rate) larger than 50-70% is required (Vanbeveren et al., 1997; Mason et al., 2001 and references therein). In section 3.4.6 we will discuss the evolution of the SN rates in the solar neighbourhood. Anticipating, to explain the observed SN Ia rate, the overall interacting binary formation rate in the intermediate mass range must be at least 40-50 %. We like to remind that in general, the binary formation rate differs from the observed binary frequency, even accounting for observational selection. A stellar population consists of evolved and non-evolved stars. Evolved stars will in most cases be observed as single stars but they could have been a secondary of an interacting binary. This secondary has become single due to a previous merger process or due to a previous SN explosion which disrupted the original binary. This means that the (observed) binary frequency in a stellar population is always smaller that the (past) binary formation rate.

First, we will present our results for a binary formation rate $f_b$ = 70% that is constant in time. The model will be referred as $f_b$1.

However, whether $f_b$ is constant in time or not is a matter of faith. The binary formation rate in regions outside the solar neighbourhood or the extra-galactic one is totally unknown. This implies that we do not know how the binary frequency varies as function of metallicity. Kobayashi et al. (1998, 2000) investigated the consequences on the time evolution of the SN Ia rates and chemical elements by introducing the Z-dependent SD supernova model. It is easy to understand that one may expect similar results from a model where the SD model is Z-independent but the binary formation rate is Z-dependent. This is the reason why we made test calculations assuming that the binary formation rate increases linearly with metallicity giving a total binary fraction $f_b$ = 0 at Z=0 and =70% at Z=0.02. The model will be referred as $f_b$2. We like to emphasise that the latter model is introduced mainly to illustrate what happens when the conservative assumption 'the binary fraction is constant in time' is abandoned.

*The standard parameter set for the galactic calculations*

Without further specification, we perform our galactic calculations assuming the standard PNS parameters listed in section 2. To describe the formation of the Galaxy (equations 3 and 5) we adopt the parameters of model A in Chiappini et al. (1997).

**3.3. The R-values**

In Brussels we have a very extended stellar evolutionary library (till the end of core helium burning) of single stars and interacting binaries covering the integration limits of the previous subsection. We like to remind that the library accounts for overshooting as it was introduced by Schaller et al. (1992), for the most recent formalisms of SW mass loss during all phases of massive star evolution, for the process of RLOF and mass transfer in case A and case $B_r$ binaries, for the CE process in case $B_c$ and case C binaries, for the CE and spiral in process in binaries with a compact companion.

Starting from this library it is straightforward to calculate the R-values of single stars. The binary tables are multidimensional and they contain the value of $R_{M1,i}$, $R_{M2,i}$ and the moments of ejection as function of primary mass ($M_1$), mass ratio (q) and orbital period (P). The details how these tables were constructed will be discussed in a separate paper (De Donder and Vanbeveren, 2001) where we will also study the binary yields. The tables are available upon request.

A few remarks are appropriate.

**a. Single stars.**

Our library contains stellar evolutionary tracks till the end of CHeB. To obtain the yields produced in the final stages we need to link our tracks to evolutionary computations that go beyond CHeB. We proceed as follow.

- $1 \leq M/M_\odot < 8$: to obtain the chemical yields in this mass range it is essential to use calculations that treat post-CHeB dredge-up phases in detail. We take the yields with the corresponding remnant masses from the work of Renzini & Voli (1981) for their case A, $\alpha$=1.5 and $\eta$=0.33. Although Renzini and Voli do not consider overshooting in their calculations, the amount of overshooting accounted for in our evolutionary computations is very moderate. Test computations show that the final results are rather robust to the introduced inconsistency.

- $8 \leq M/M_\odot \leq 100$: we calculated the evolution of single stars till the end of CHeB accounting for the most recent SW formalisms and we relate our models with the nucleosynthesis computations of WW95 (case B SN model) to estimate the corresponding SN yields. This relation is based on the final CO-core masses at the end of CHeB.

- We do not consider stars with masses lower than solar, however they are taken into account in our chemical evolutionary simulations where they are important for locking up matter; we start the integration from 0.1 $M_\odot$.

**b. Interacting binary stars.**

- During RLOF (case B) the mass loser loses layers that are CNO processed and thus mainly enriched in $^4$He and $^{14}$N. When the RLOF is non-conservative and matter lost by the primary also leaves the binary we explicitly account for the interstellar enrichment.

- In case of massive primary stars we proceed similarly as for single stars i.e. we link the CO core at the end of our CHeB calculations to the nucleosynthesis results of WW95 (case B SN) to compute the SN yields.

- Intermediate mass primaries lose all their hydrogen rich layers and most of the helium layers outside the helium burning shell by RLOF. The remnant mass equals the mass of the final CO core. This remnant will become a WD. Remark that the CHeB and post CHeB evolution of an intermediate mass primary is substantially different from the evolution of a single star with the same mass. An important difference is the CO-enrichment of the ISM which is not expected from the binary components.

- The evolution of the mass gainer is followed in detail. When the gainer has a compact companion, its further evolution is governed by the RLOF and/or CE phase. Some of these systems are SN Ia candidates. However, most of them will merge (see next point).

- In the present work mergers are treated as follows:

- for the merging of two MS stars we assume that the mass of the merger is equal to the total mass of the binary at the moment of contact. It is very likely that its structure is similar to that of a MS star that has been rejuvenated by accretion. For simplicity we take the yields of a single star of corresponding mass.
- when a Thorne-Zytkow object is formed with a NS in its core, we adopt the model of Cannon et al. (1992) according to which the R(S)G is completely evaporated by an efficient SW and the compact star reappears. In case of a BH in the centre, we assume that the R(S)G is completely swallowed without ejecting material.

- PNS predicts that many WDs with a MS/RG companion will merge. As argued earlier in section 2, they may produce Type II SNe. The physics of the latter is still very uncertain but even more uncertain is the chemical composition of the ejected matter if a SN happens. Therefore, in the following computations we have not included the contribution from merged WD+MS/RG systems.

- The chemical composition of the SNIa ejecta likely depends on the nature of the progenitor systems (i.e. the WD's initial mass, initial C/O value, the accretion rate, etc.) and may be different for the SD and DD model. However we do not focus on this point and use the updated version (Iwamoto et al., 1999) of the classical W7 nucleosynthesis yields (Nomoto et al.,1984; Thielemann et al., 1986) which are at present still in good agreement with the observed spectra of a typical SNIa. Our $R_M$ tables explicitly account for these SN Ia yields since for any given intermediate mass binary our PNS code decides whether an SN Ia will happen or not.

- We do not account for the enrichment by nova outbursts. Since they are not potential producers of the here considered isotopic elements (but rather of isotopes like $^{15}$N, $^{17}$O etc.) their absence will only marginally affect our results.

- Low mass binaries ($M_{1,0} < 3\ M_\odot$) enrich the ISM in He. The binary yields are very similar to the low mass single star yields and, therefore, in equation 6 we treat all low mass stars as single stars.

**3.4. Model Results.**

As mentioned earlier, the effects of all binary parameters on the chemical yields will be discussed separately in De Donder and Vanbeveren (2001). Here we concentrate on overall effects on GEMs.

To illustrate the effects of binaries, the following models are selected:

| Model | Combination |
|---|---|
| a | Standard PNS parameters - SD model with a Z-independent WD wind - $f_b 2$ |
| b | Similar to model a but with $f_b 1$ |
| c | Standard PNS parameters – SD model with a Z-dependent WD wind - $f_b 2$ |
| d | Similar to model c but with $f_b 1$ |
| e | Standard PNS parameters - DD model with a flat q-distribution - $f_b 2$ |
| f | Similar to model e but with $f_b 1$ |
| g | Standard PNS parameters - DD model with a Hog. q-distribution - $f_b 2$ |
| h | Similar to model g but with $f_b 1$ |
| i | Standard PNS parameters - only single stars with SNIa binary progenitors which are followed in detail according to the SD model with a Z-independent WD wind and for $f_b 2$ |
| j | Similar to model i but with $f_b 1$ |
| k | Model a but with the WW95 Fe yields reduced by a factor of two |
| l | Model b but with the WW95 Fe yields reduced by a factor of two |
| m | Model c but with the WW95 Fe yields reduced by a factor of two |
| n | Model d but with the WW95 Fe yields reduced by a factor of two |
| o | Model g but with the WW95 Fe yields reduced by a factor of two |
| p | Model h but with the WW95 Fe yields reduced by a factor of two |
| q | Model k with the addition that stars initial more massive than $40 M_\odot$ eject 4 $M_\odot$ before a BH is formed |

Models i and j deserve some attention. We like to compare the results of a galactic model where all binaries are included in detail with a simplified (unrealistic) model where only the binaries are included that produce SN Ia events and all other stars are considered as single.

### 3.4.1. The overall star formation rate (SFR).

The computed star formation rate shown in figure 3 is typical for the adopted two-infall galactic model. A first phase of high activity and a second phase of moderately continuous star formation. Our model predicts a present star formation rate $\Psi(M_\odot/pc^2/Gyr)$ in the solar neighbourhood of ~2.54 that lies within the observed range of (2-10) $(M_\odot/pc^2/Gyr)$ (Güsten & Mezger, 1982). The corresponding present gas density $\sigma_g(M_\odot/pc^2)$ estimated by our model is 6.98, giving a gas fraction of 14% and a total surface mass density ($\sigma_T$) of ~50 $(M_\odot/pc^2)$. Observations for the solar neighbourhood are $\sigma_g = 6.6 \pm 2.5$ and $\sigma_g/\sigma_T = (0.05-0.20)$ (Rana, 1991). Although interacting binaries return less matter to the ISM because of a higher formation rate of NSs and BHs, our computations reveal that this effect is negligible in the overall SFR even for a high constant binary formation rate $f_b$=70%. So we conclude that

- *the rate of star formation only marginally depends on whether binaries form or not.*

**Figure 3:** The theoretical predicted star formation rate (for $f_b2$) as function of time in the solar neighbourhood. The star-formation-stops resulting from the threshold in the surface gas density are skipped.

### 3.4.2. Age-metallicity relation

The age-metallicity relation (AMR) shows the time evolution of the ratio [Fe/H] which is usually taken as a measure of the metallicity of the Galaxy. Because of the observed nonnegligible scatter in metallicity at all ages there is currently some uncertainty on whether an AMR exists in the disk. The scatter is suggested to come from inhomogeneous chemical evolution which is an effect that we cannot account for since our model assumes instantaneous mixing of recycled gas and a homogeneous steady infall and star formation. Therefore we look only at the average relation. The AMR predicted by different models together with the observed data from Edvardsson et al. (1993) is shown in figures 4a and 4b. Our AMR differs from the AMR of other groups. This is due mainly to the massive star yields used in the galactic code which depend significantly on the SW mass loss formalisms during the various evolutionary phases, on the formalism to treat convective core overshooting, on BH formation in massive stars and, last not least, on the effect of binaries and the link with SN Ia's (remind that our SN Ia's do not occur earlier than ~1 Gyr after the beginning of the formation of the Galaxy).

Most of our models predict an AMR where Fe increases too fast during the early phases of galaxy evolution compared to the observational trend (provided that the observations are not object of major errors). In these early phases only massive stars dictate the variation of [Fe/H]. The degree of increment is very sensitive to the massive star iron SN ejecta (especially to those of stars with M>20M$_\odot$) and the latter depend critically on the physics of the SN explosion, in particular on the adopted explosion energy and the 'mass cut'. To illustrate, the case A SN model of WW95 predicts massive star iron yields that are lower by about a factor of two compared to the case B yields. Figure 4a shows the effect (the models k, m, o) and we conclude that during the early phases of galactic evolution, the case A SN model of WW95 gives an overall better fit to the averaged observed trend of data. Adjusting the Galactic model parameters like the inflow amplitude and accretion timescales may also slow down the growth of the iron abundance, however a discussion of the effect of these parameters is beyond the scope of the present work

After ~1 Gyr, intermediate mass stars start to eject iron through Type Ia SNe. Their relative contribution to the total production of iron, depends primarily on the SNIa progenitor population, mass ratio distribution, binary evolutionary parameters, and the binary formation rate. We find that about 30% of the present total iron contents is made by SNIa's when the SD model is used and ~50% when the DD model applies in combination with a flat q-distribution. In the latter case, the sudden increase in [Fe/H] that appears between 5 and 6 Gyr is due to a maximum in the formation rate of SNIa's showing up at ~4 Gyr (see figure 1a subsection 2.3).

Figures 4a and 4b illustrate that

- *the AMR predicted by a galactic code where the SN Ia SD model is Z-dependent and the SN Ia progenitor binary frequency is Z-independent are similar to those where the SD is Z-independent but the binary frequency is Z-dependent.*

When we compare the AMR of models i and j with the ones of models a and b we can conclude that

- *Except for the effect of SN Ia's, the AMR predicted by a theoretical galactic model hardly depends on whether binaries are included or not.*

**Figure 4a:** The theoretically predicted AMR in the solar neighbourhood for the models with $f_b2$. The bold points represent the observational data (Edvardsson et al., 1993).

**Figure 4b**: The same as figure 4a but for the models with $f_b1$.

### 3.4.3. G-dwarf metallicity distribution

Because of their long main-sequence lifetime (~10-15 Gyr) G dwarfs ($0.8 \leq M/M_\odot \leq 1.05$) provide spectra that correspond with the chemical composition of the interstellar gas at early phases of the Galaxy. Their distribution as function of metallicity reflects the chemical evolutionary history of the ISM. Figure 5 shows the G dwarf distribution in the solar neighbourhood predicted by our galactic code for different models. A comparison with the observational derived distributions of Wyse & Gilmore (1995) and of Rocha-Pinto & Maciel (1996) immediately rules out a number of combinations. As a general conclusion, during the early phases of galactic evolution, the case A SN model corresponds better to the observations than the case B SN model, however, when the intermediate mass binary frequency is large and constant in time, the case A SN model must be combined with a Z-dependent SN Ia rate to find a reasonable fit. The figure also illustrates that a model with only single stars but with a detailed treatment of the SN Ia progenitors predicts a very similar G-dwarf distribution as a galactic model where the effects of all binaries are included. And finally, the figure forces us to conclude that the predicted G-dwarf distribution depends critically on the adopted model for the SN Ia progenitors. Since the parameters describing the formation of a galaxy and the star formation history are estimated by comparing the predicted and observed G-dwarf distributions, also the estimated values will depend on the adopted SN Ia progenitor model.

**Figure 5:** The theoretically predicted G dwarf metallicity distribution in the solar neighbourhood for different models. The observationally derived histograms are from Wyse & Gilmore (1995) (dashed line) and from Rocha-Pinto & Maciel (1996) (full line).

### 3.4.4. Solar abundances

In figure 6 (a,b,c) we show the predicted mass fractions, normalised to the observed solar abundances, of the elements in the ISM at the time and place of the birth of the Sun (=10.5 Gyr). As can be noticed all our model predictions reproduce the observations within a factor of two. When direct BH formation for single stars already starts at 25 $M_\odot$ (BH1), the metal abundances are reduced since the final CO core is swallowed by the BH. Mainly O, Ne and Mg are underabundant by a factor 2. We conclude that

- *The observed abundances are best reproduced when direct BH formation occurs above 40 $M_\odot$ for single and binary components (BH2).*

The overabundance of carbon in some models is always due to the intermediate mass stars. Since the carbon yields in intermediate mass interacting binaries is significantly smaller than in intermediate mass single stars, an obvious way to reduce the carbon discrepancy is to increase the binary formation rate in the intermediate mass range (see also subsection 3.4.5), i.e.

- *The observed solar carbon abundance may be indirect evidence for the presence of a significant intermediate mass binary population during the whole evolution of the solar neighbourhood.*

Similarly as for the AMR, we performed test calculations with a galactic model where, except for the SN Ia's, all binaries were omitted and all stars were treated as single stars. The results are included in figure 6c. We conclude that

- *Except for carbon and iron, a model with single stars only but with a correct treatment of the SN Ia population predicts similar solar abundances as a model where the effects of all binaries are included in detail.*

**Figure 6a:** The theoretical predicted solar abundances (normalised to the observed solar values from Anders & Grevesse (1989)) for the different cases of BH formation i.e. BH1 (filled squares) and BH2 (open triangles) both in combination with the SNIa SD model with a Z-independent WD wind and with $f_b2$.

**Figure 6b:** Similar as figure 6a. but for model e (open triangles) and model g (filled squares).

**Figure 6c:** Similar as figure 6a. but for model b (open triangles) and model i (filled squares).

### 3.4.5. Abundance ratios evolution

Figures 7 to 13 display the time behaviour in terms of metallicity, of the abundance ratios of the various elements ($^{12}C$, $^{14}N$, $^{16}O$, $^{24}Mg$, $^{28}Si$, $^{32}S$ and $^{40}Ca$) with respect to iron. Our predicted ratios are normalised to the OBSERVED solar ratio. To compare, observational data from different sources (indicated in the legend) are plotted.

*Carbon and the intermediate mass binary connection*

All galactic models where in the early phases of galactic evolution, the intermediate mass binary formation rate is low, predict a significant [C/Fe] bump at [Fe/H] ~ -0.5. However, observations (Laird, 1985; Tomkin et al., 1986; Carbon et al., 1987) do not show this behaviour, i.e. [C/Fe] remains more or less constant around 0. Nucleosynthesis flaws in intermediate mass single stars has been suggested to explain the inconsistency and a process called 'hot bottom burning' (HBB) was invented (see for example Samland, 1998). However, as illustrated below, binaries offer a solution as well and may be HBB is not really necessary.

Iron is produced by explosive burning in all SNe, also in SNIa's if the overall intermediate mass binary formation rate is large. Carbon is made as a primary element during CHeB and is ejected by stars over the whole massive star range via a SN explosion and/or a stellar wind whereas, in the intermediate mass range, primarily single stars enrich the ISM in carbon. The carbon production from interacting intermediate mass primaries is smaller than the single star counterparts because thermal pulses are mostly avoided due to previous mass transfer whereby most of the H envelope is lost. Only in systems that undergo late case C RLOF (i.e. during the TP-AGB phase) dredge-up of $^{12}$C from deeper layers can occur to enrich the outer layers. However, these systems have a very low formation rate (compared to the other cases). Therefore, it can be expected that the enrichment of [C/Fe] caused by intermediate mass stars during disk evolution is smaller when binaries are included. The foregoing explains the relatively large effect of binaries on the time evolution of carbon and may provide a solution for the [C/Fe]-bump discrepancy. Our galactic simulations reveal that the [C/Fe]-bump is largely reduced in models where the intermediate mass binary formation is high already during the early evolutionary phases of the solar neighbourhood, i.e.

- *the observational behaviour of the ratio [C/Fe] as function of [Fe/H] in the solar neighbourhood can be explained if the intermediate mass binary formation was high during the whole evolution of this galactic region.*

Interestingly, combining the A-M relation, the G-dwarf distribution and the carbon variation, we are inclined to conclude that the following model gives the better correspondence with observations: constant high binary frequency (at least in the intermediate mass range), case A SN model of WW95, Z-dependent SNIa rates as predicted by the SD scenario.

*A binary solution for the enrichment of nitrogen during the early evolutionary phases of a galaxy*

Despite the large scatter, observations show that [N/Fe] remains 0 down to [Fe/H]=-3 indicating the production of primary nitrogen over the whole metallicity range. Primary nitrogen production by massive stars at low metallicity is not predicted by current evolutionary models of massive stars which is a well-known problem in CEMs.

The effects of rotation on stellar evolution was studied in detail by Maeder and Meynet (2000a, b) and could provide a solution. There may also be a massive binary solution for the formation of primary nitrogen during the early evolutionary phases of a galaxy. In a close binary with a massive primary, the secondary star may be polluted with carbon via accretion of SW material of the primary during core helium burning (a WC star) and/or accretion of carbon enriched matter ejected during the SN explosion of the primary. Test calculations show that through thermohaline mixing the envelope of the accretion star may become strongly enriched in carbon that will be synthesised partly into nitrogen during H shell burning. The foregoing scenario was also suggested by Vanbeveren (1994) to explain OBC type stars. Detailed stellar evolutionary computations where this effect is included have not yet been made, but we like to present the following suggestion:

- *the enrichment of primary nitrogen during the early evolutionary phases of a galaxy may be caused by massive close binary interaction.*

*Oxygen, Magnesium, Silicon, Sulphur and Calcium and the connection with the massive black hole formation*

Massive stars are the primary producers of oxygen and dominate the [O/Fe] evolution during the early phases of the Galaxy. Early observations of oxygen abundances in dwarfs and G and K giants suggested that [O/Fe] is nearly constant (~0.5) for [Fe/H] ≤ -1 (e.g. Gratton et al., 1986) (i.e. during halo evolution) and then falls gradually (the so-called knick) to the solar value due to the high iron and low oxygen production by Type Ia SNe. Since the time of appearance of Type Ia SNe depends on the evolutionary time scale of the binary progenitors, the knick has often been considered as a constraint on the SNIa progenitor model. More recent observational work (Israelian et al., 1998; Boesgaard et al., 1999) show that oxygen is overabundant in the galactic halo with an increase of [O/Fe] from 0.6 to 1 going from [Fe/H]=-1 to -3. With these new data the downturn to solar value is much less pronounced and almost absent.

Our model predictions with the standard galactic model give an overall underproduction of oxygen relative to iron during the early galactic evolution. The downturn to solar value appears at values of [Fe/H] higher than -1 (on average at ~ -0.3) corresponding with the maximum formation rate of Type Ia SNe as predicted by the progenitor models. When the DD model combined with a flat q-distribution is used, the decrease is stronger because, although in both models the first SNIa's appear at more or less the same moment, more SNIa's are formed on a shorter time scale especially in the age-interval [0.15-0.9] Gyr (see fig. 1.a).

To obtain a better fit during the early galactic evolutionary phases we reduced the massive star Fe yields with a factor of two as we did earlier to obtain a better AMR. Although the results are better, we still find an underproduction of oxygen during the halo evolution. A higher production of oxygen can be achieved if stars with an initial mass above $40M_\odot$ do undergo a SN explosion instead of directly collapsing into a BH as we assumed in our standard model for massive stars. It has been suggested by Nelemans et al. (1999) that to explain the runaway status of the BH X ray binaries Cyg X-1 and Nova Sco 1994, a SN explosion with ejection of $\sim(3-4)M_\odot$ should have occurred prior to BH formation indicating that oxygen ejection in the mass range $\geq 40M_\odot$ may not be negligible. To illustrate this effect we assumed that all stars above $40M_\odot$ eject $2M_\odot$ of oxygen and $2M_\odot$ of carbon. It is clear from figure 9 (model q) that the correspondence with the observations at low metallicities is considerably improved. So we conclude here that despite their low formation rate as predicted by the adopted IMF, high mass stars that form massive BHs with a forgoing SN explosion are important for the oxygen enrichment of the ISM during the early evolutionary phases of the galactic evolution, i.e.

- *the variation of the ratio [O/Fe] as function of [Fe/H] during the early phases of galactic evolution may be an indirect indication that the formation of ALL massive black holes is preceded by a SN explosion where significant amounts of oxygen is ejected into the ISM.*

Our results for the evolution of the other α elements relative to iron are given in figures 10 to 13. We find an overall evolution similar as for oxygen. The evolutionary behaviour is in general agreement with the average observed trend though the ratios are rather low, especially for magnesium and sulphur, at low metallicity whether binaries are included or not. The same arguments as for the oxygen discrepancy may also apply here although uncertainties in the yields of these elements listed in WW95 may be important as well (Chiappini et al., 1999).

*Comparison with other computations*

The oxygen-discrepancy during the early galactic evolution was also found by Timmes et al. (1995) who also use the WW95 case B stellar models as we do. When rescaling the figures of Chiappini et al. (1997) relative to the OBSERVED solar abundance, the discrepancy is also visible in their results. At first glance it was surprising that this discrepancy was not found by some other groups (e.g. Chiappini et al, 1999; Boissier and Pranzos, 2000; Prantzos and Boissier, 2000; Portinari et al., 1998; Samland, 1998). Although most groups use the WW95 yields, it is not always clear whether the case A or case B SN results are implemented in the galactic code. Portinari et al. (1998) use the case A yields which corresponds to our Fe/2 model. The WW95 library provides yields for stars with an initial mass lower than 40 $M_\odot$. The authors argue that stars with initial mass larger than 40 $M_\odot$ undergo direct BH formation without a SN explosion, similarly as is assumed in the present paper. However, at least some groups mentioned above neglect the BH-argument and EXTRAPOLLATE the WW95 yields for stars with initial mass larger than 40 $M_\odot$. In the overall IMF, this mass range is very insignificant, but during the early phases of galactic evolution, mainly the massive stars play a role and among the massive stars, the M ≥ 40 $M_\odot$ subclass plays a very significant role as far as the oxygen enrichment is concerned. Actually, what they did seems rather arbitrary but after second thought it resembles what we did in order to solve the oxygen discrepancy, i.e. although we allow for the formation of massive BHs in the mass range M ≥ 40 $M_\odot$ (their existence is indicated by observations of X-ray binaries), we let them precede by a SN explosion where mass layers with limited mass are ejected into the ISM [supported by recent observations of the standard high mass X-ray binary Cyg X-1 (see Nelemans et al., 1999)].

***Overall conclusion resulting from the evolution of the abundance ratios:***

Similarly as done for the AMR and for the predicted solar abundances, we also made test computation with a galactic model where, except for the details of the SN Ia progenitors, we consider all other stars as single. We conclude that

- *the time evolution of the elements $^4$He, $^{16}$O, $^{20}$Ne, $^{24}$Mg, $^{28}$Si, $^{32}$S and $^{40}$Ca predicted by galactic chemical evolution where the effect of binaries is considered in detail differ by no more than a factor of two to three from the results of models where most of the stars are treated as single stars and where the effect of binaries is simulated only to account for the SNIa population.*

**Figure 7 a&b:** The theoretically predicted evolution of [C/Fe] as function of [Fe/H]. The observational data (filled circles) are taken from Laird (1985), Tomkin et al. (1986), Gratton & Ortolani (1986) and Carbon et al. (1987).

**Figure 8 a&b:** The theoretically predicted evolution of [N/Fe] as function of [Fe/H]. The observational data are taken from Laird (1985), Gratton & Ortolani (1986) and Carbon et al. (1987).

**Figure 9 a&b:** The theoretically predicted evolution of [O/Fe] as function of [Fe/H]. The observational data are taken Gratton et al. (1996) and Boesgaard et al. (1999).

**Figure 10 a&b:** The theoretically predicted evolution of [Mg/Fe] as function of [Fe/H]. The observational data are taken from Gratton & Sneden (1987), Magain (1987, 1989) and Edvardsson et al. (1993).

**Figure 11 a&b:** The theoretically predicted evolution of [Si/Fe] as function of [Fe/H]. The observational data are taken from Tomkin et al. (1984), François (1986), Gratton & Sneden (1991) and Edvardsson et al. (1993).

**Figure 12 a&b:** The theoretically predicted evolution of [S/Fe] as function of [Fe/H]. The observational data are taken from Clegg et al. (1981) and François (1987, 1988).

**Figure 13 a&b:** The theoretically predicted evolution of [Ca/Fe] as function of [Fe/H]. The observational data are taken from Tomkin et al. (1985), Gratton & Sneden (1988, 1991) and Edvardsson et al. (1993).

### 3.4.6. Supernova Rates.

Supernova statistics are severely affected by uncertainties inherent to observations and coming from small number statistics. From a large set of SN data recent estimates of the local rate of SNe have been derived by Cappellaro et al. (1997, 1999). If it is assumed that the Galaxy is of Type Sbc with a total luminosity $L_B=2\times10^{10}$ $L_\odot$ they predict a total galactic rate of ~2 SNe per century of which ~17% SNIa's, 12% SNIb,c's and 71% SNII's.

In our evolutionary model SNIb,c's come from massive binary components that have lost their H envelope through SW and/or RLOF and from massive single stars that become WR stars. We do not consider the explosion of a coalesced WD with its MS or RG companion star, although they could be potential SNII producers (see subsection 2.4). Figure 14 gives the evolution with time of the SN rate per century for the different types as predicted with the SD model combined with the standard galactic evolutionary model (section 3.2). We find a present distribution of 66% SNII's, 20% SNIbc's and 14% SN Ia's giving a total rate of ~1.4 SNe per century, that approaches the observed values. At very low metallicity mainly Type II SNe are formed since all massive single stars explode as a Type II and also because of the low binary formation rate. With increasing metallicity more SNIb,c's form when the binary frequency increases and massive single stars start to loose their H envelope by a SW. As shown in figure 15a, the ratio SNII/SNIb,c quickly evolves to a constant value ~3 after the formation of the halo-thick disk. Accounting for the special class of SN II's discussed in section 2, this ratio could be ~4.

The SNIa rate relative to the total SN rate increases along with the binary formation rate to a present value of 14% which is also the observed ratio. With the SD model, the present SN rates corresponds to the single star and binary formation history of the last 1-2 Gyr. This explains why the predicted present SN rates only marginally depend on whether the Z-dependent or the Z-independent SD scenario is adopted or whether a variable or constant binary formation rate applies.

We like to emphasise that the present SN Ia rate relative to the total present SN rate depends linearly on the binary formation rate at the moment the progenitors were formed. For the SD model the progenitors were formed 1-2 Gyr ago when the (intermediate mass) binary formation rate was 70%. Therefore, we conclude that

- *when SN Ia's are formed via the SD model, to obtain an SN Ia rate as observed in the solar neighbourhood an intermediate mass binary formation rate $\geq$ 70% is required.*

Hachisu et al. (1999b) proposed a wider symbiotic channel to Type Ia SNe. In their model a significant fraction of the binaries with initial period larger than 10 years could become WD + RG binaries which may produce SN Ia's. They estimate that about 1/3 of all the SN Ia's could have such an origin. It is straightforward to estimate the consequence on the conclusion made above on the intermediate mass binary formation rate $f_b$ as we defined it in section 3.2, i.e. the value '70%' is reduced to ~40-50%.

Because of the variety in binary evolutionary channels through which SNe of all type can be produced, it is clear that galactic evolution of the absolute and relative SN rates depends on the PNS parameters, binary formation rate and the adopted progenitor model for SNIa's. This is demonstrated in figure 15b where we show the results for other models. It can be concluded that for the DD model with a Hogeveen mass ratio distribution we need an intermediate mass binary formation rate larger than 50%, however in combination with a flat distribution a lower interacting binary frequency (~20-30%) suffices to meet the observations.

The behaviour of the SN rates at low metallicity depends on the behaviour of the binary formation rate and the SW mass loss with changing metallicity. Supernova studies at high red shift will shed more light on this however it will be difficult to disentangle the predictions of a model with a Z-dependent SN Ia model and a Z-independent binary formation rate from a model with a Z-independent SN Ia model and a Z-dependent binary formation rate.

**Figure 14:** The predicted SN rate (per century) as function of time for the different types i.e. SNII (thin line), long dashed line SNIb,c (thick full line) and SN Ia (grey line). Model a is used.

**Figure 15a:** The evolution in time of the ratio SNII/SNIb,c computed with model a.

**Figure 15b:** The evolution of the SNIa rate relative to the total SN rate for the models a (full black line), c (short dashed black line), e (short dashed grey line) and g (full grey line).

**4. Summary and Conclusions.**

In this paper we have studied the time evolution of several isotopes in the solar neighbourhood with a two infall galactic model. We accounted for the formation of binaries and followed their evolution in detail.

Our simulations demonstrate that galactic chemical evolutionary models that account in detail for the evolution of close binaries predict a time-variation of the chemical elements $^4$He, $^{16}$O, $^{20}$Ne, $^{24}$Mg, $^{28}$Si, $^{32}$S and $^{40}$Ca that differ by no more than a factor 2-3 from the evolutionary results predicted by models that only account for single star evolution and the formation of Type Ia SNe. Obviously, binaries are very important for iron and they may be very important to understand the time evolution of carbon and nitrogen. By comparing prediction and observation of the age-metallicity relation, the G-dwarf distribution and the time variation of carbon, we are tempted to conclude that the best correspondence is achieved by a CEM that assumes a large (intermediate mass) binary frequency, a Z-dependent single degenerate WD scenario for the SN Ia rate and the case A SN models of WW95, at least when Z is small during the early phases of galactic evolution.

Most interesting, from our chemical simulations we derived some constraints on the stellar evolution of massive stars. The details of the BH formation process have an important impact on the time evolution of the α elements. The solar O, Ne and Mg abundances are only well reproduced when ALL massive stars explode even the stars more massive than $40 M_\odot$ that form massive BHs.

Finally, we followed the evolution of the Galactic SN rate of the Types II, $I_{b,c}$ and Ia. In the early phases of galaxy evolution, the total SN rate is dominated by the SNII's. After about 1.5 Gyr, the ratio $SNII/SNI_{b,c}$ becomes more or less constant with a value of ~3-4 which is in agreement with the presently observed ratio. Assuming that Type Ia SNe arise from exploding CO WDs of (super)Chandrasekhar mass, we recover the present SNIa/SN(Ia+Ibc+II) ratio of ~20%.

We like to end with a general comment: accounting for the uncertainties in chemical evolutionary codes that contain a considerable number of (unknown and thus free) parameters, the uncertainties in overall binary statistics, the uncertainties in the physics of late stages of stellar evolution, the uncertainties in stellar evolution in general, the uncertainties in observations, when one finds correspondence between theoretical prediction and observations one may wonder what is the meaning. At present it looks like a bottomless pitfall and therefore, before we will have really reliable galactic models, much more research is necessary in single star and binary astrophysics.

**Acknowledgement**

We thank the referee. Prof. Dr. Bernard Pagel, for his critical and very valuable remarks. We sended preprints of the present paper to many colleagues and we hereby thank everybody for her/his suggestions. We received pro's and contra's and it soon became clear that there was a correlation with understanding and not-understanding the effects of binaries and binary evolution. Therefore, when reading the present manuscript, please contact us if something is not clear.

**References.**

Abt, H.A., 1983, ARA&A., 21, 343.

Anders, E., Grevesse, N., 1989, Geochim. Cosmochim. Acta, 53, 197.

Boesgaard, A. M., King, J. R., Deliyannis, C. P., Vogt, S. S., 1999, AJ., 118, 2542.

Boissier, S., Prantzos, N., 2000, MNRAS, 312, , 398.

Cannon, R. C., Eggleton, P.P., Zytkow, A. N., Podsiadlowski, P., 1992, ApJ., 386, 206.

Cappellaro, E., Turatto, M., Tsvetkov, D. Yu., Bartunov, O.S., Pollas, C., Evans, R., Hamuy, M., 1997, A&A., 322, 431.

Cappellaro, E., Evans, R., Turatto, M., 1999, A&A., 351.

Carbon, D.F., Barbuy, B., Kraft, R.P., Friel, E.D., Suntzeff, N.B., 1987, PASP, 99, 335.

Carlberg, R. G., Dawson, P. C., Hsu, T., Vandenberg, D. A., 1985, ApJ., 294, 674.

Chiappini, C., Matteucci, F., Gratton, R., 1997, ApJ., 477.

Chiappini, C., Matteucci, F., Beers, T. C., Nomoto, K., 1999, AJ., 515, 226.

Chiosi, C., 1980, A&A., 83, 206.

De Donder, E., Vanbeveren, D., 1998, A&A., 333, 557.

De Donder, E., Vanbeveren, D., 1999, New Astronomy, 4, 167.

De Donder, E., Vanbeveren, D., 2001, submitted

Edvardsson, B., Andersen, J., Gustafsson, B., Lambert, D. L., Nissen, P. E., Tomkin, J., 1993, A&A, 275, 101.

Fryer, C., 1999, ApJ., 522, 413.

Fujimoto, M.Y., Ikeda, Y., Iben, I. Jr., 2000, ApJ., 529, L25.

Garmany, C.D., Conti, P.S., Massey, P., 1980, ApJ, 242, 1063.

Gratton, R., Carretta, E., Matteucci, F., Sneden, C., 1996, in The Formation of the Galactic Halo, ASP Conference Series, Vol.92, eds. H. Morrison and A. Sarajedini.

Greggio, L., Renzini, A., 1983, A&A, 118, 217.

Güsten, R., Mezger, P.G., 1982, Vistas Astron., 26, 3.

Hachisu, I., Kato, M., Nomoto, K., 1996, ApJ., 470, L97.

Hachisu, I., Kato, M., Nomoto, K., Umeda, H., 1999a, ApJ., 519, 314.

Hachisu, I., Kato, M., Nomoto, K.: 1999b, ApJ 522, 487.

Hogeveen, S.J., 1991, Ph.D. Thesis, University of Amsterdam.

Iben, I., Jr., Tutukov, A., 1984, ApJSS, 54, 335.

Israelian, G., Garcia Lopez, R. J., Rebolo, R., 1998, ApJ., 507, 805.

Israelian, G., Rebolo, R., et al., 1999, Nature, 401, 142.

Iwamoto, K., Brachwitz, F. et al., 1999, ApJSS., 125, 439.

King, A.R., Pringle, J.E., Wickramasinghe, D.T., 2001, MNRAS, 320, L45.

Kippenhahn, R., Ruschenplatt, G., Thomas, H.C., 1980, A&A., 91, 175.

Kobayashi, C., Tsujimoto, T., Nomoto, K., Hachisu, I., Kato, M., 1998, ApJ., 503, L155.

Kobayashi, C., Tsujimoto, T., Nomoto, K., 2000, ApJ., 539, 26.

Laird, J.B., 1985, ApJ., 289, 556.

Landau, L.D., Lifshitz, E.M., 1959, Classical Theory of Fields, Pergamon Press.

Li, X.-D., van den Heuvel, E.P.J., 1997, A&A., 322, L9.

Lorimer, D.R., Bailes, M., Harrison, P.A., 1997, MNRAS, 289, 592.

Lubow, S.H., Shu, F.H.: 1975, ApJ. 198, 383.

Maeder, A., 1992, A&A., 264.

Maeder, A., Meynet, G., 2000a, ARAA, 48 (in press)

Maeder, A., Meynet, G., 2000b, A.A. 361, 159.

Marigo, P., Bressan, A., Chiosi, C., 1996, A&A., 313, 545.

Marigo, P., Bressan, A., Chiosi, C., 1998, A&A., 331, 564.

Mason, B. D., Hartkopf, W.I., Holdenried, E. R., Rafferty, T. J., 2001, AJ., 121, 3224.

Meusinger, H., Reimann, H. G., Stecklum, B., 1991, A&A., 245, 57.

Nelemans, G., Tauris, T.M., van den Heuvel, E.P.J., 1999, A&A., 352, L87-L90.

Neo, S., Miyaji, S., Nomoto, K., Sugimoto, D.: 1977, Publ. Astron. Soc. Japan 29, 249.

Nomoto, K., 1982, ApJ., 253, 798.

Nomoto, K., Thielemann, F.K., Yokoi, K., 1984, ApJ., 286, 644.

Pagel, B.E.J., 1980 in Normal Galaxies, Nato Course, August, Cambridge.

Pagel, B.E.J., Portinari, L., 1998, MNRAS., 298, 747.

Popova, E.I., Tutukov, A.V., Yungelson, L.R., 1982, Astron. Space Sci., 88, 55.

Portinari, L., Chiosi, C., Bressan, A., 1998, A&A., 334.

Prantzos, N., Boissier, S., 2000, MNRAS, 313, , 338.

Rana, N., 1991, ARA&A., 29, 129.

Renzini, A., Voli, M., 1981, A&A., 94, 175.

Rocha-Pinto, H.J., Maciel, W.J., 1996, MNRAS, 279, 447.

Rossi, S., Beers, T.C., Sneden, C., 1999, ASP Conf.Ser., 165, 264.

Salpeter, E.E., 1955, ApJ., 121, 161.

Samland, M., 1998, ApJ., 496, 155.

Scalo, J.M., 1986, Fund. of Cosm. Phys., 11, 1.

Schaller, G., Shaerer, D., Meynet, G., Maeder, A., 1992, A&AS., 96, 269.

Schmidt, M., 1959, ApJ., 129, 243.

Schmidt, M., 1963, ApJ., 137, 758.

Segretain, L., Chabrier, G., Mochkovitch, R., 1997, ApJ., 481, 355.

Sparks, W.M., Stecher, T.P., 1974, ApJ., 188, 149.

Talbot, R.J., Arnett, D.W., 1975, ApJ., 197, 551.

Thielemann, F.K., Nomoto, K., Yokoi, Y.: 1986, A&A., 158, 17.

Timmes, F. X., Woosley, S. E., Weaver, T. A., 1995, ApJ.SS., 98.

Tinsley, B.M., 1980, Fund. Cosmic Phys., 5, 287.

Tomkin, J., Sneden, C., Lambert, D. L., 1986, ApJ., 302, 415.

Vanbeveren, D., Herrero, A., Kunze, D., Van Kerkwijk, M., 1994, Space Sci. Rev. 66, 395.

Vanbeveren, D.: 1994, Space Sci. Rev. 66, 327.

Vanbeveren, D., De Loore, C., 1994, A&A., 290, 129.

Vanbeveren, D., 1995, A&A., 294, 107.

Vanbeveren, D., Van Bever, J., De Donder, E., 1997, A&A., 317, 487-502.

Vanbeveren, D., De Donder, E., Van Bever, J., Van Rensbergen, W., De Loore, C., 1998a, New Astronomy, 3, 443.

Vanbeveren, D., De Loore, C., Van Rensbergen, W., 1998b, A&Arv., 9, 63.

Vanbeveren, D., 2000, in The Evolution of the Milky Way, p. 139., Kluwer Academic Publishers, eds. Matteucci, F.and Giovannelli, F.

Vanbeveren, D., 2001, in The Influence of Binaries on Stellar Population Studies, p. 249., Kluwer Academic Publishers, eds. D. Vanbeveren.

Webbink, R.F., 1984, ApJ., 277, 355.

Whelan, J., Iben, I., 1973, ApJ., 186, 1007.

Wyse, R.F.G., Gilmore, G., 1995, AJ., 110, 2771.

Woosley, S. E., Weaver, T. A., 1995, ApJ.SS., 101.

Yoshii, Y., Tsujimoto, T., Nomoto, K., 1996, ApJ., 462, 266.

______________________________

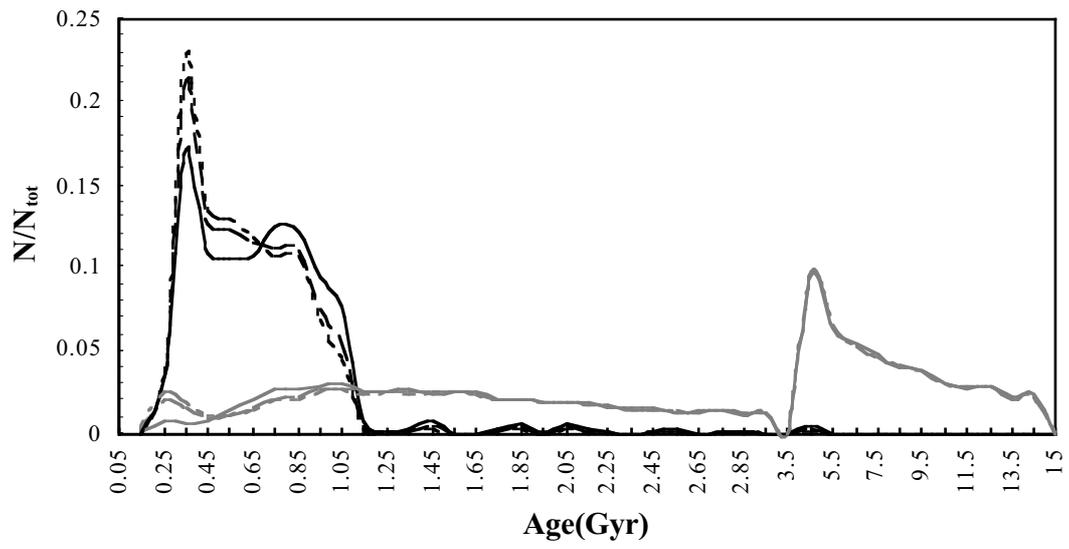

Figure 1a.

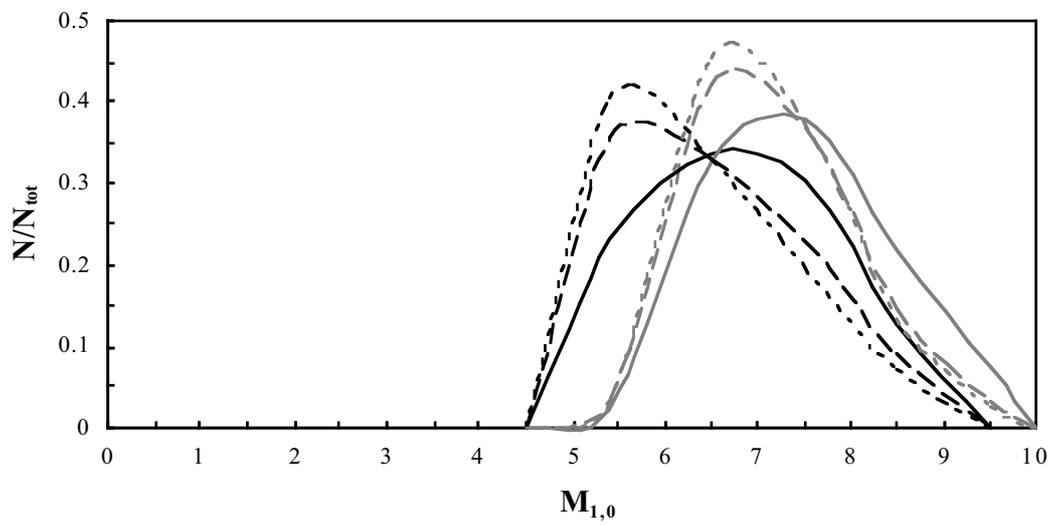

Figure 1b.

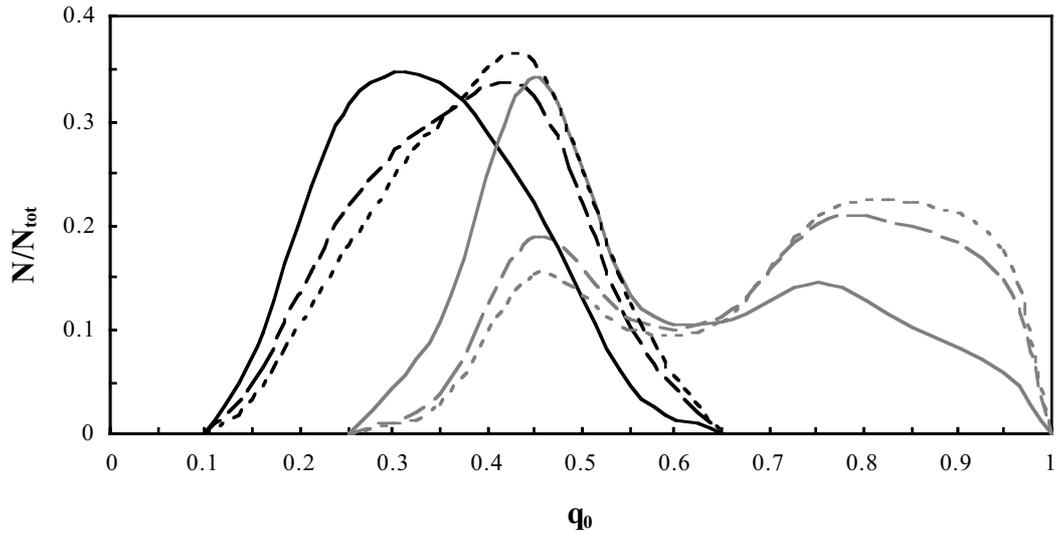

Figure 1c.

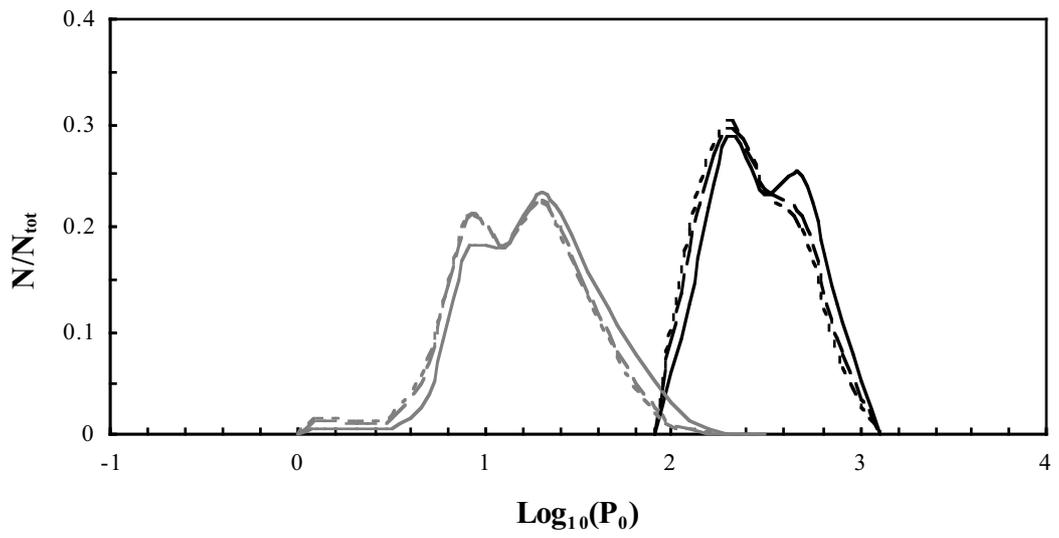

Figure 1d.

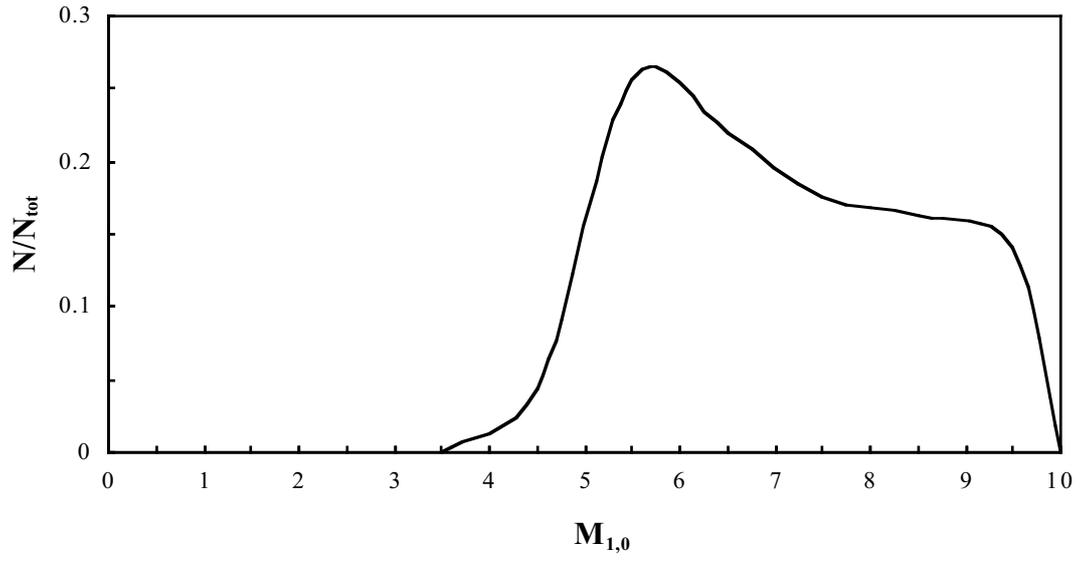

Figure 2a.

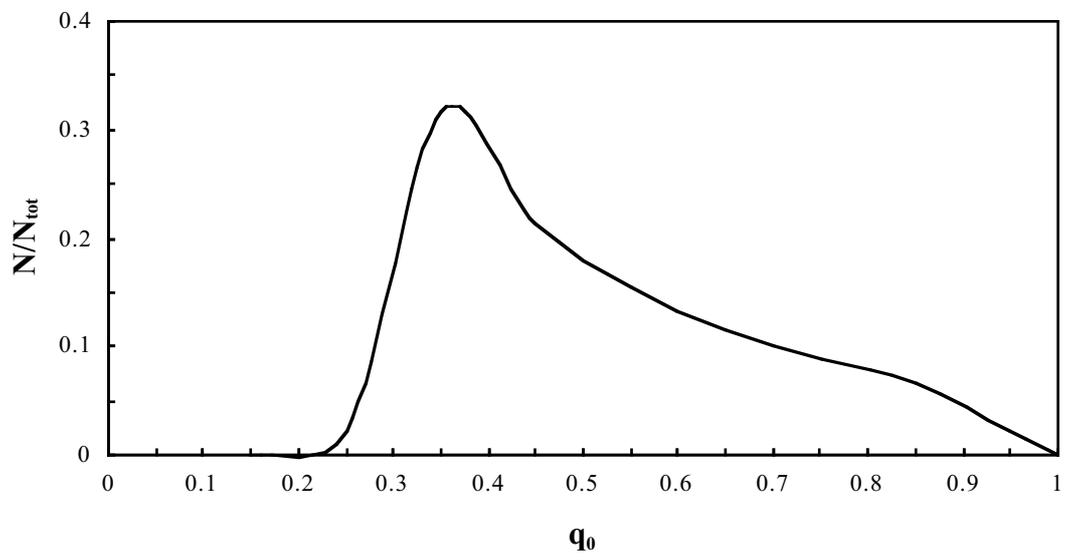

Figure 2b.

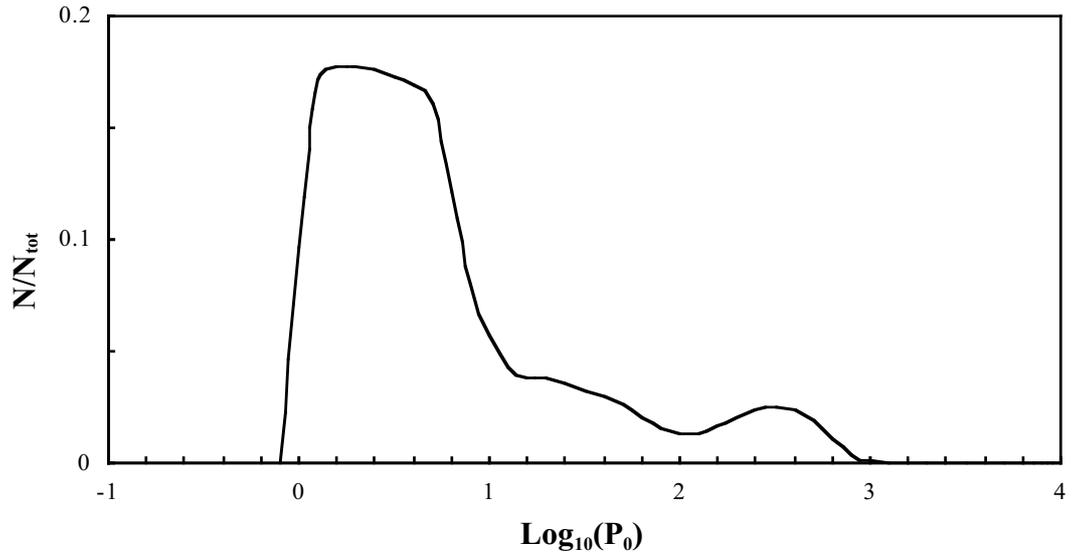

Figure 2c.

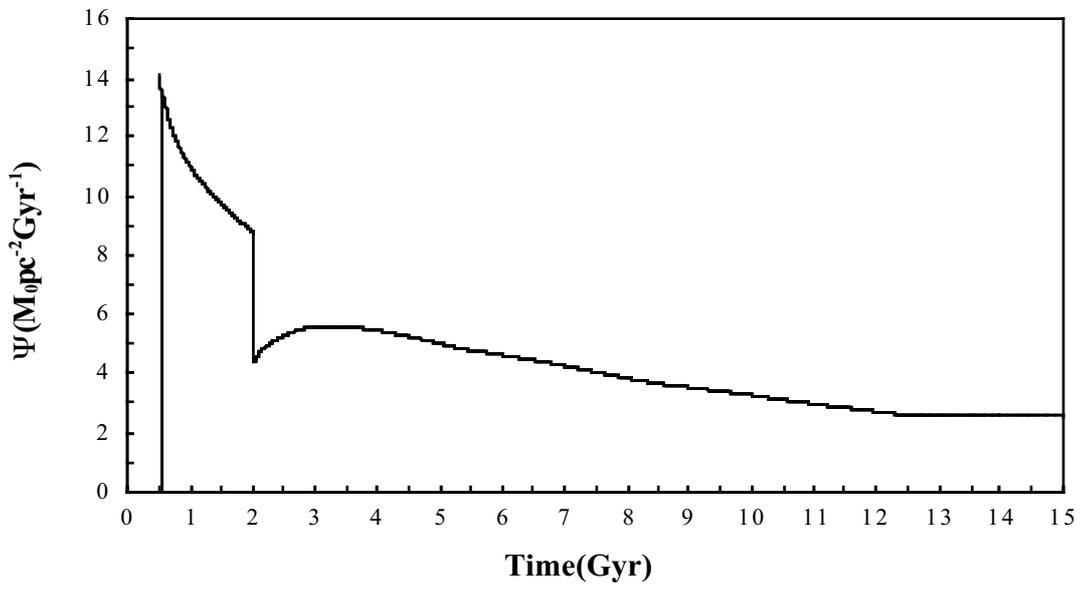

Figure 3.

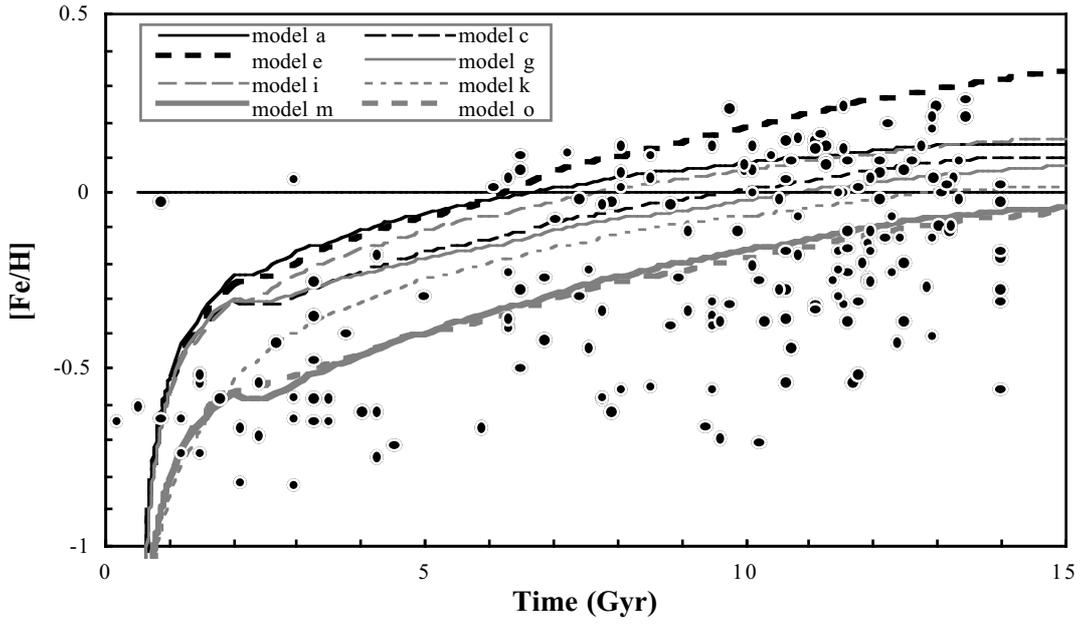

Figure 4a.

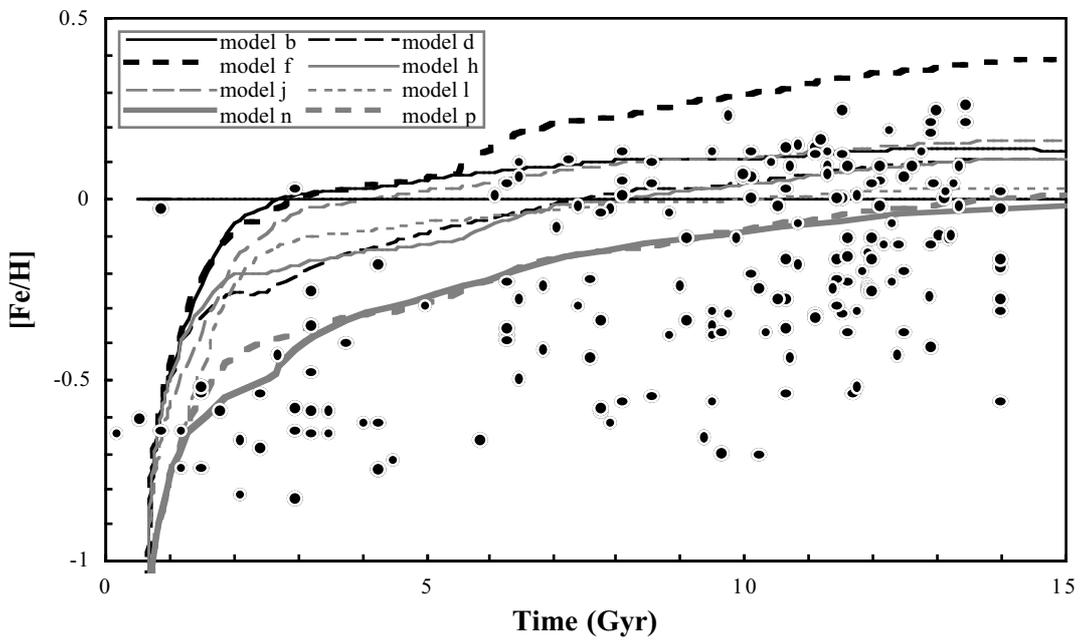

Figure 4b

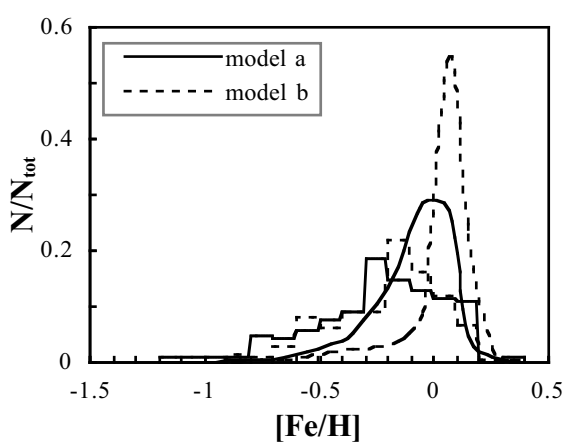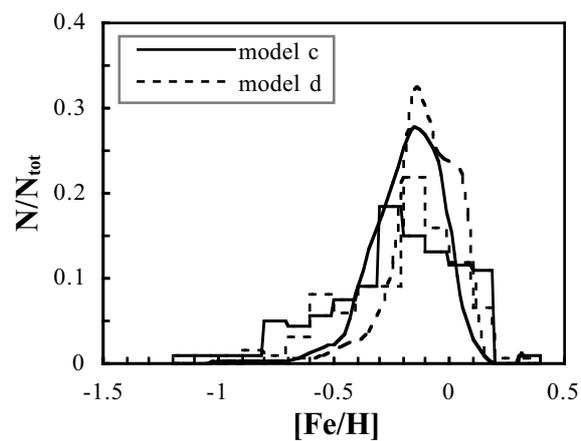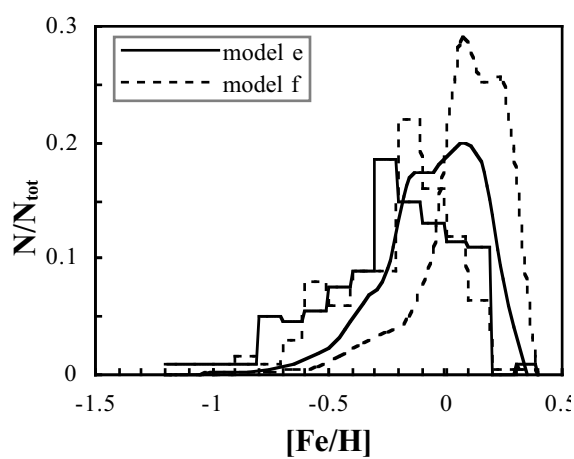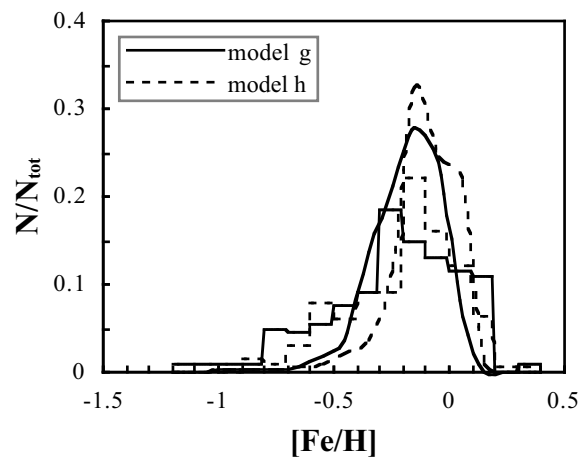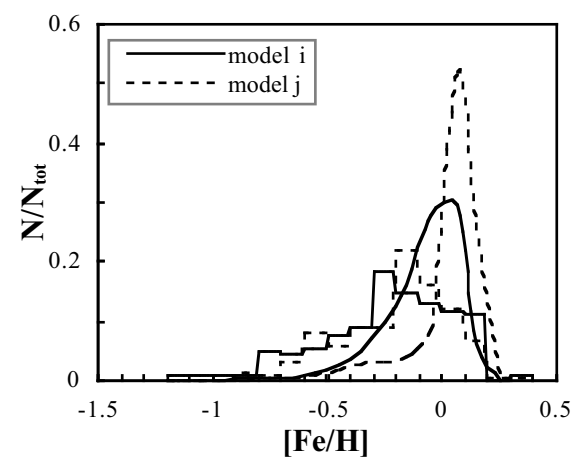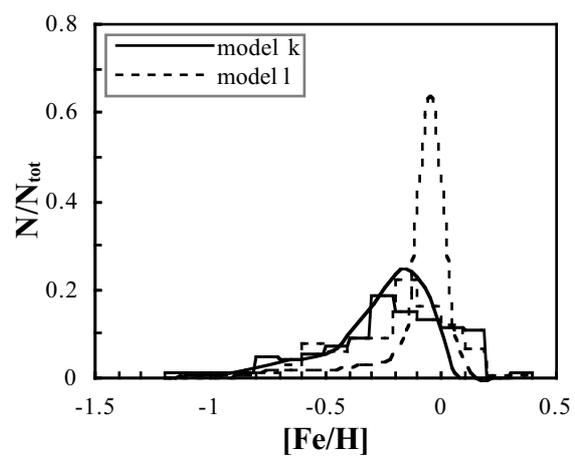

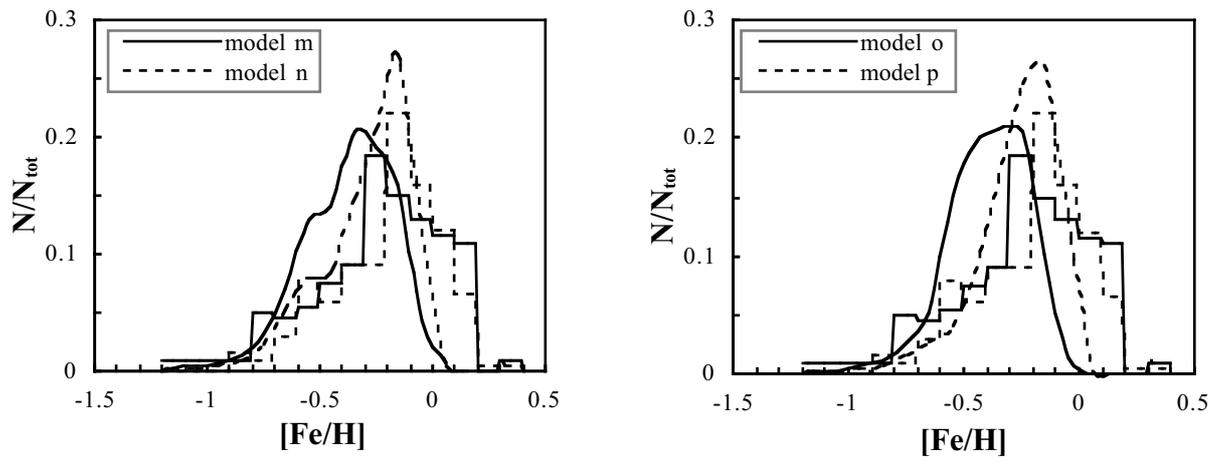

Figure 5.

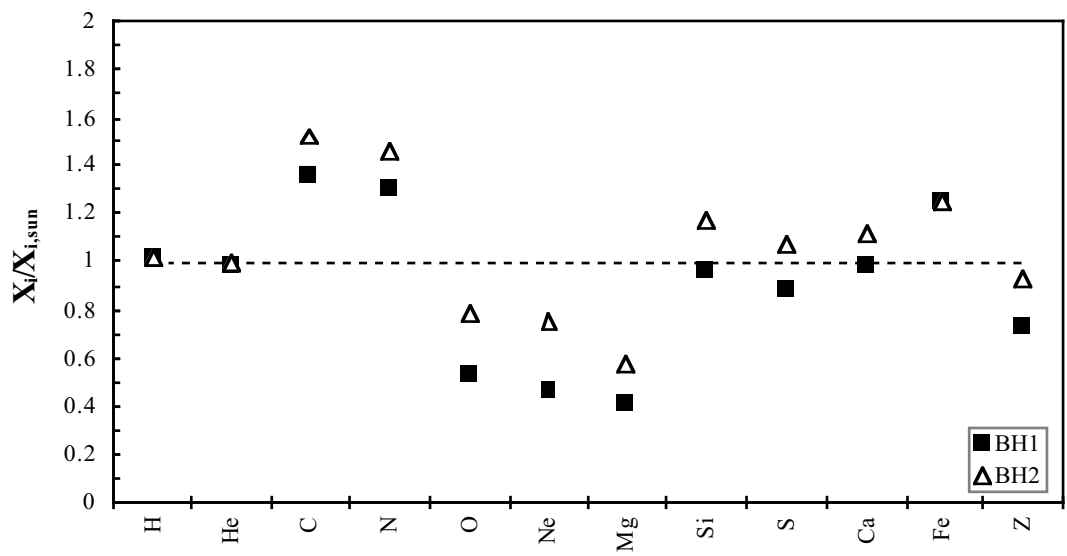

Figure 6a.

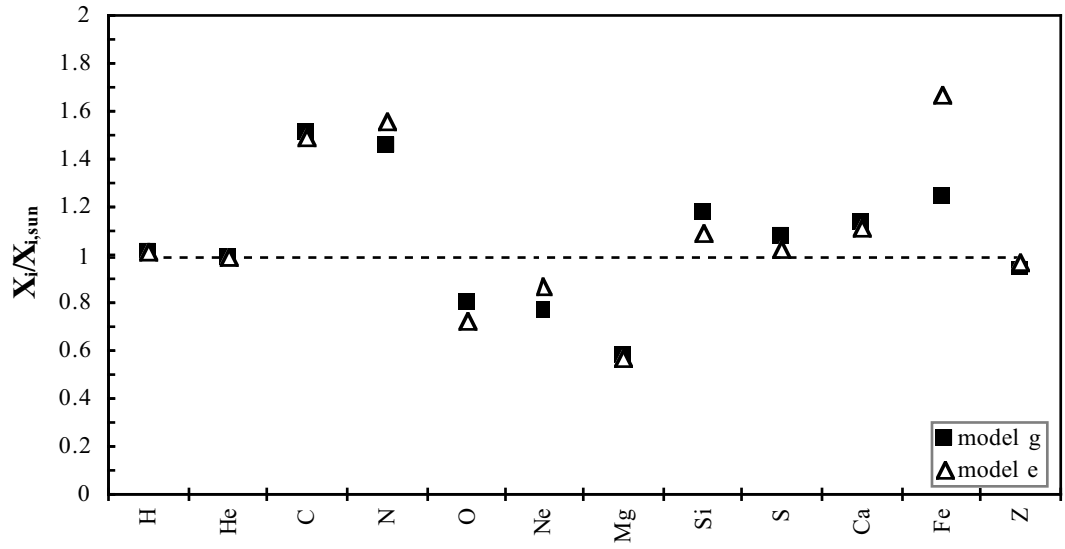

Figure 6b.

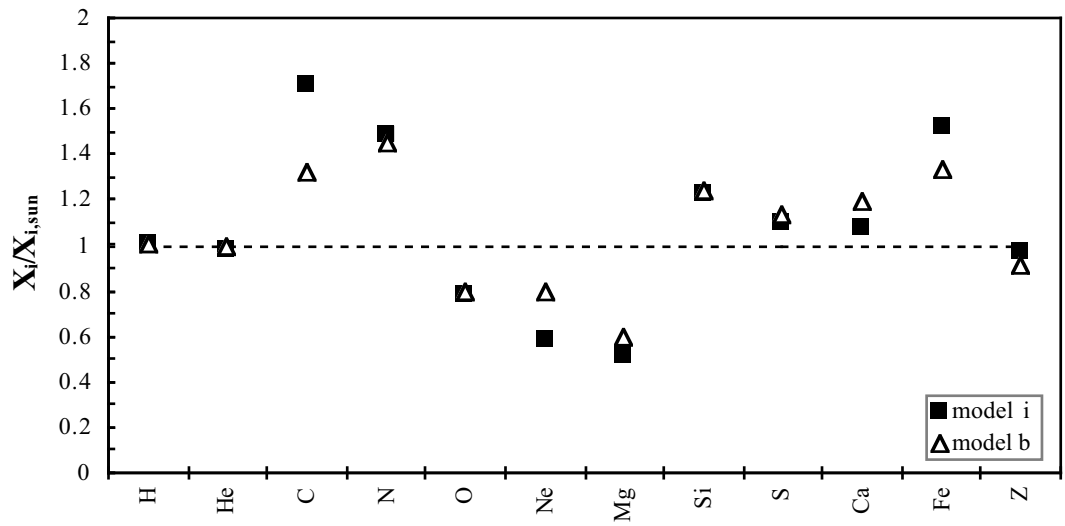

Figure 6c.

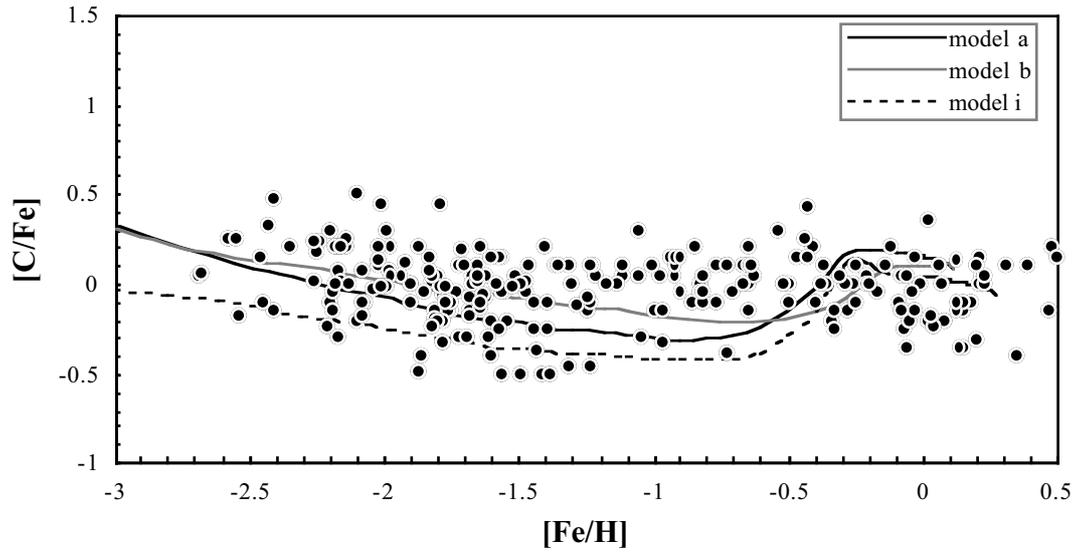

Figure 7a.

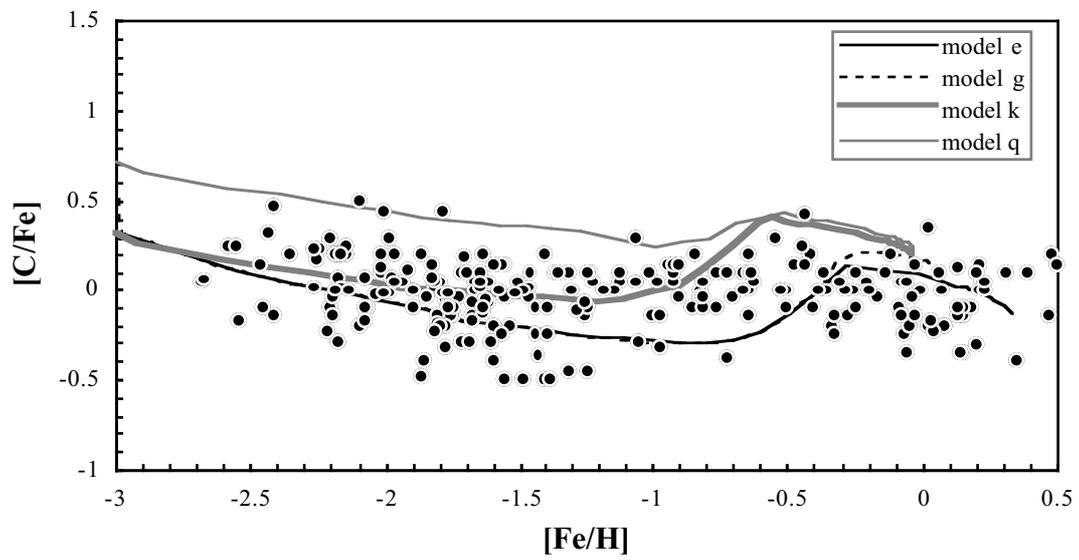

Figure 7b.

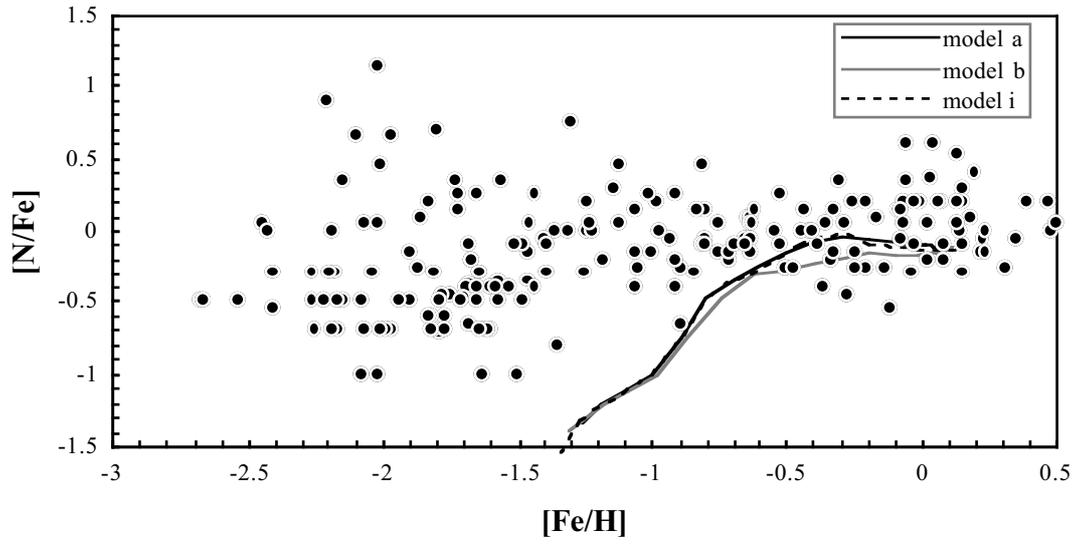

Figure 8a.

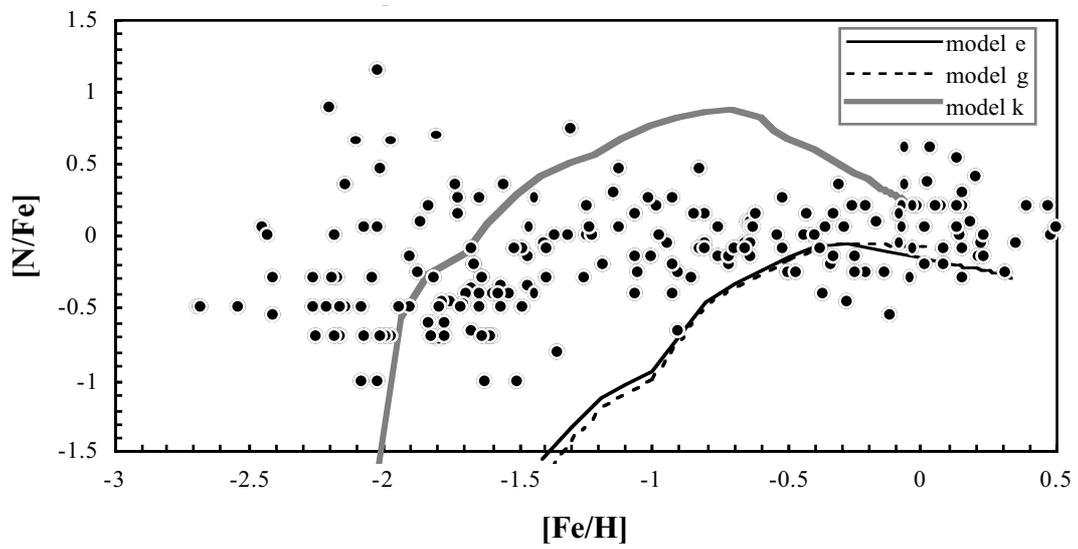

Figure 8b.

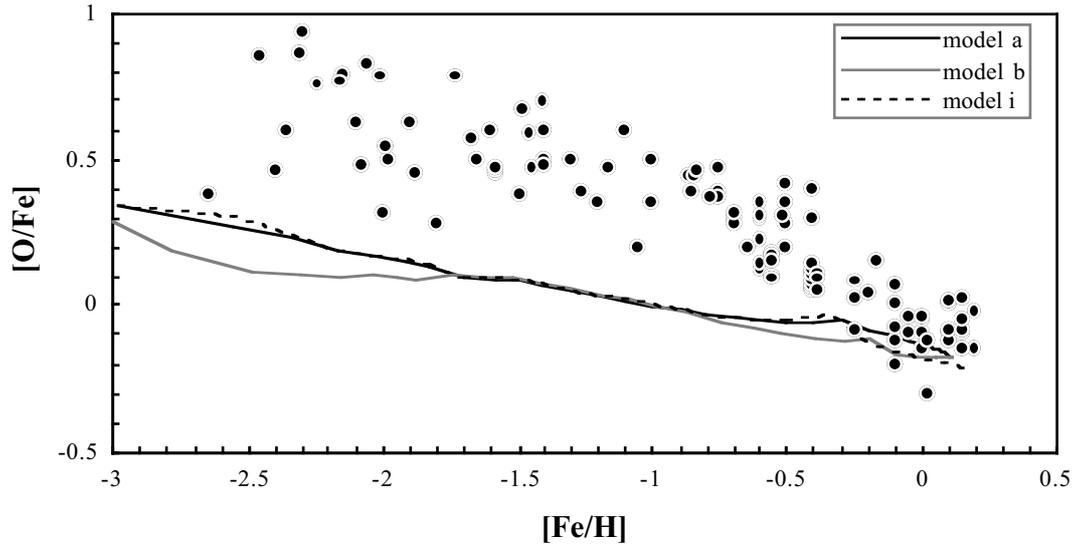

Figure 9a.

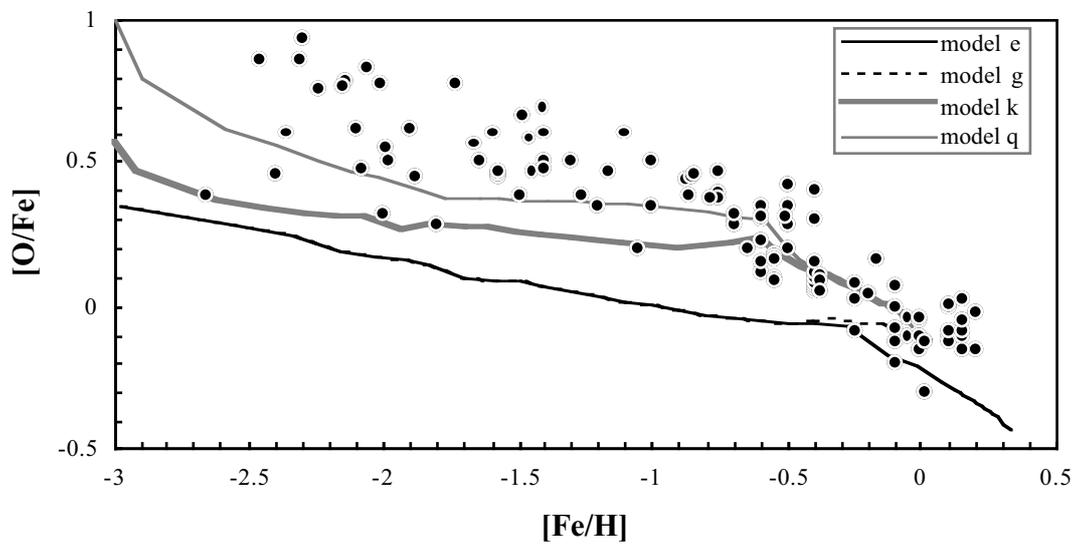

Figure 9b.

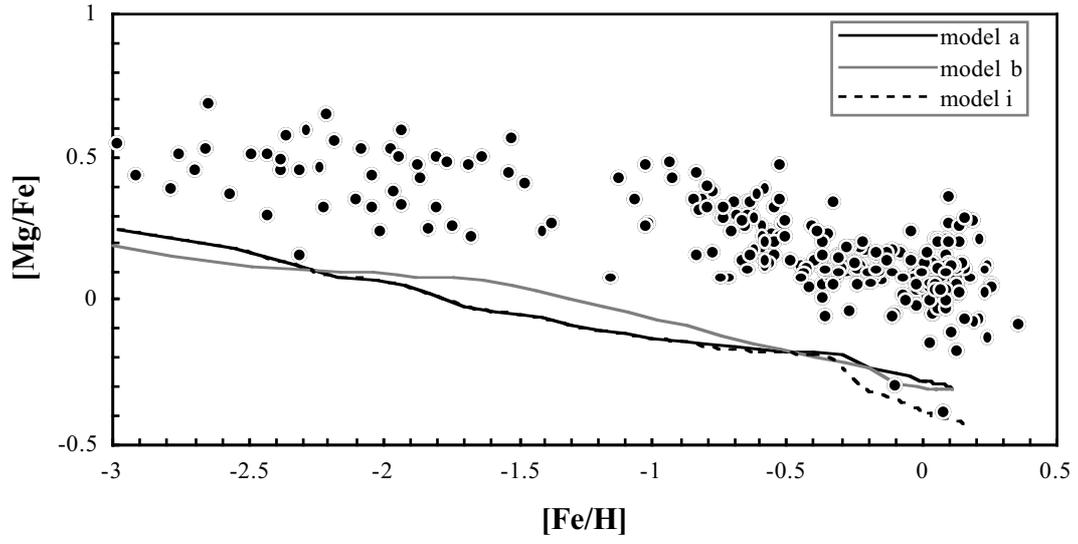

Figure 10a.

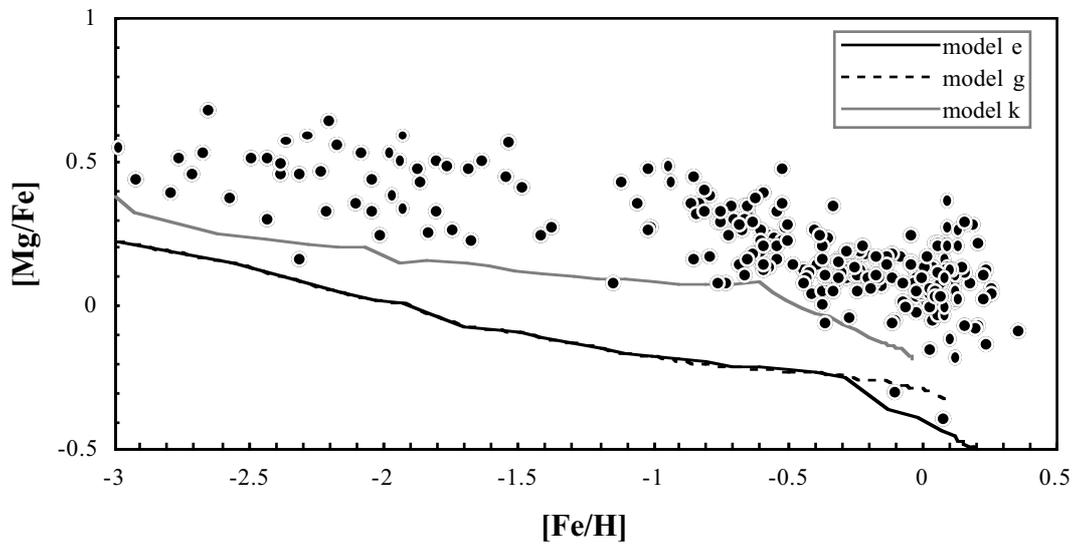

Figure 10b.

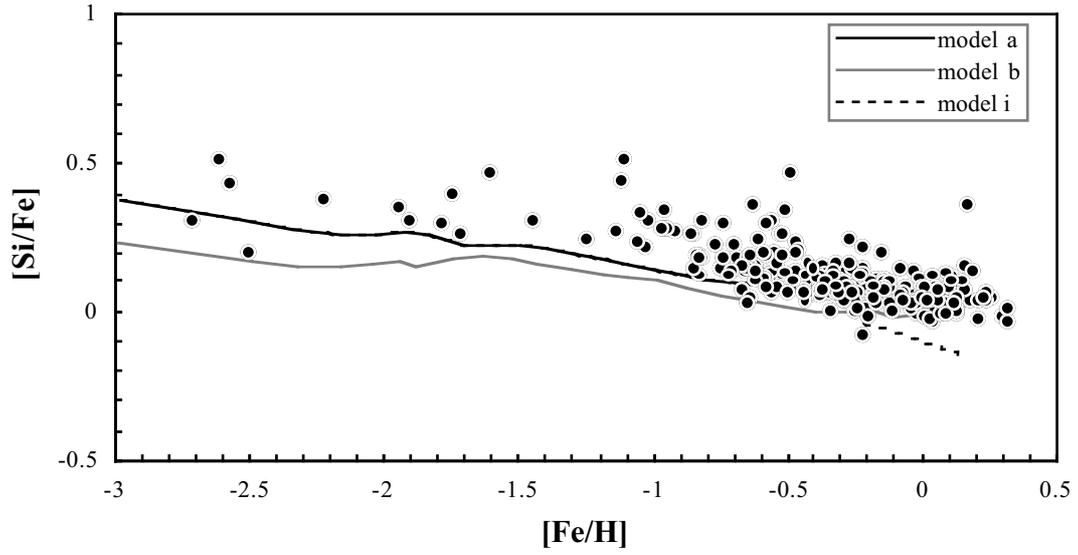

Figure 11a.

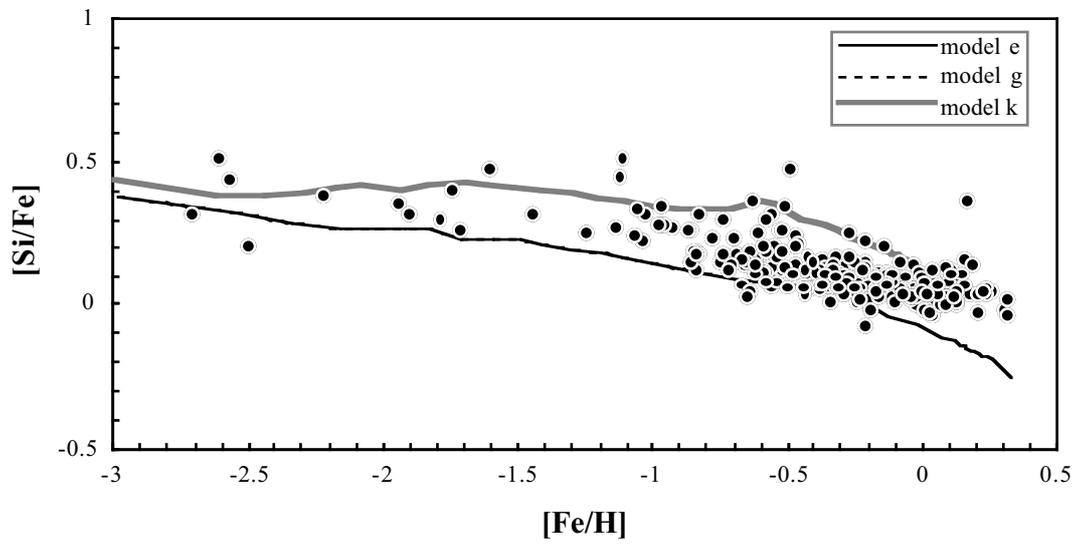

Figure 11b.

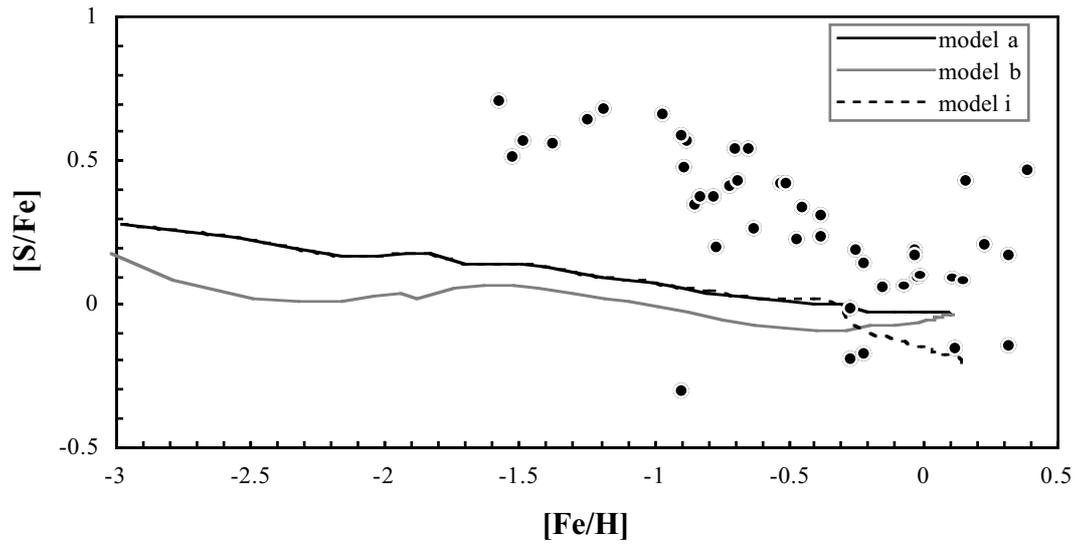

Figure 12a.

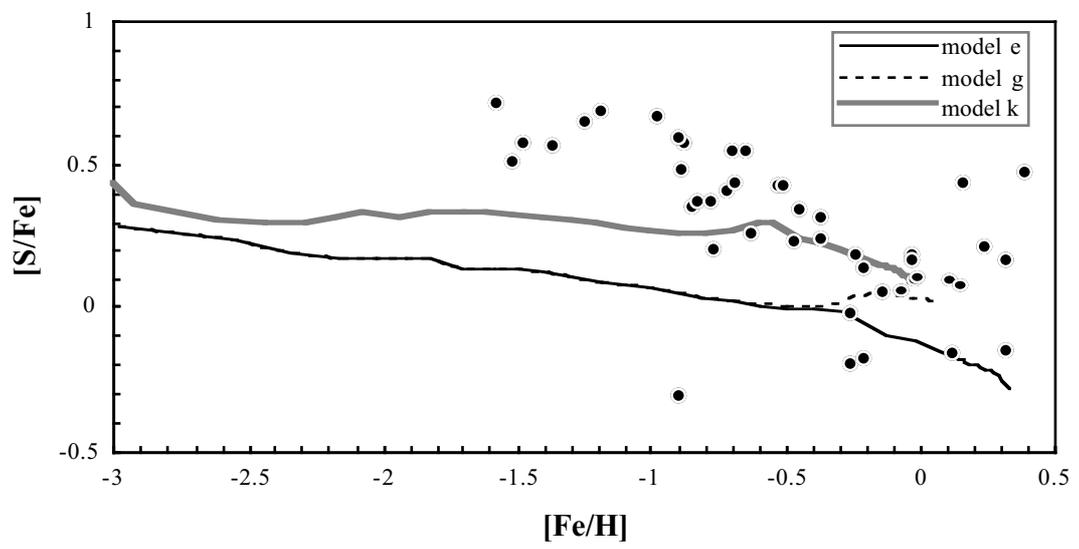

Figure 12b.

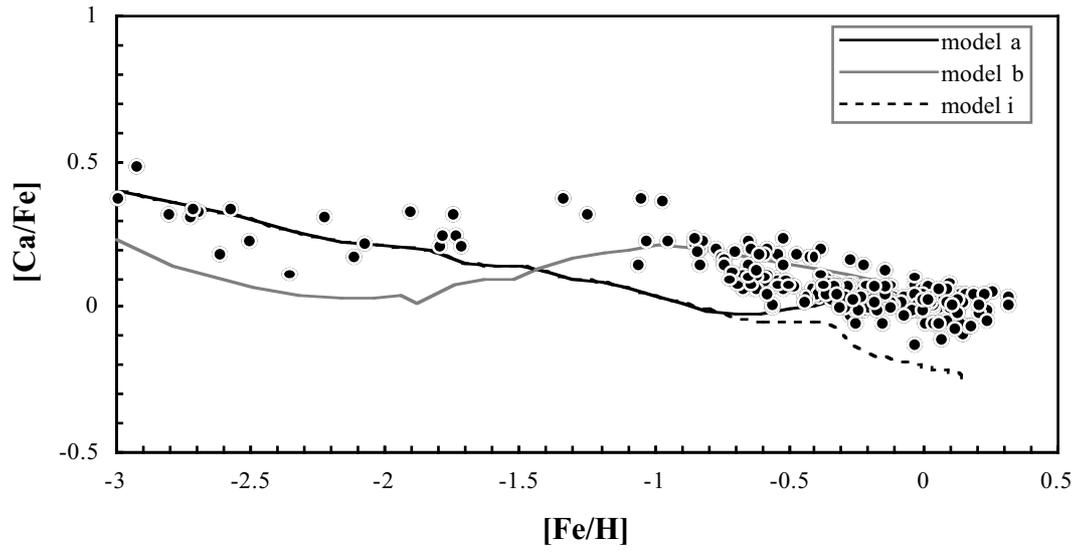

Figure 13a.

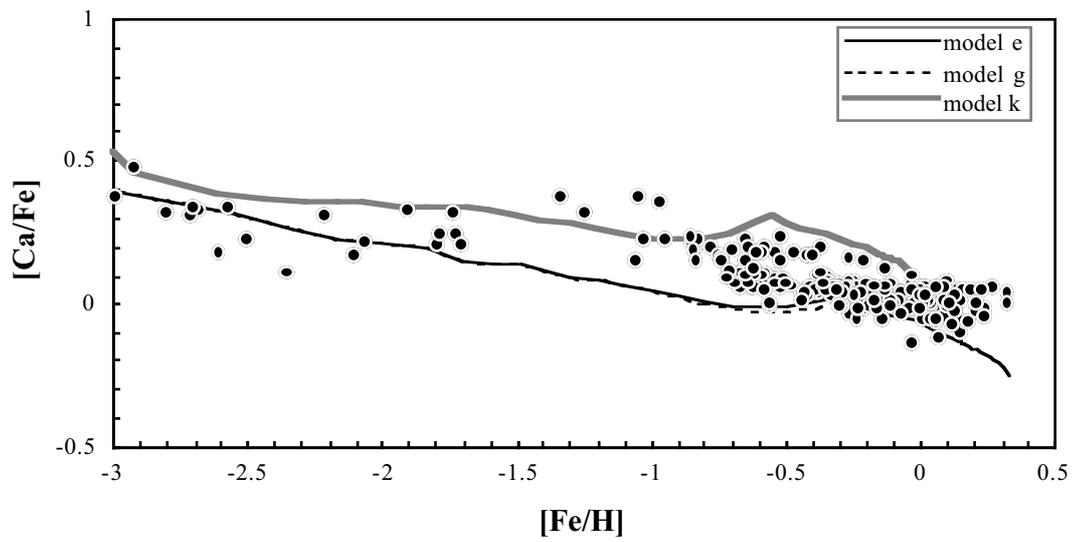

Figure 13b.

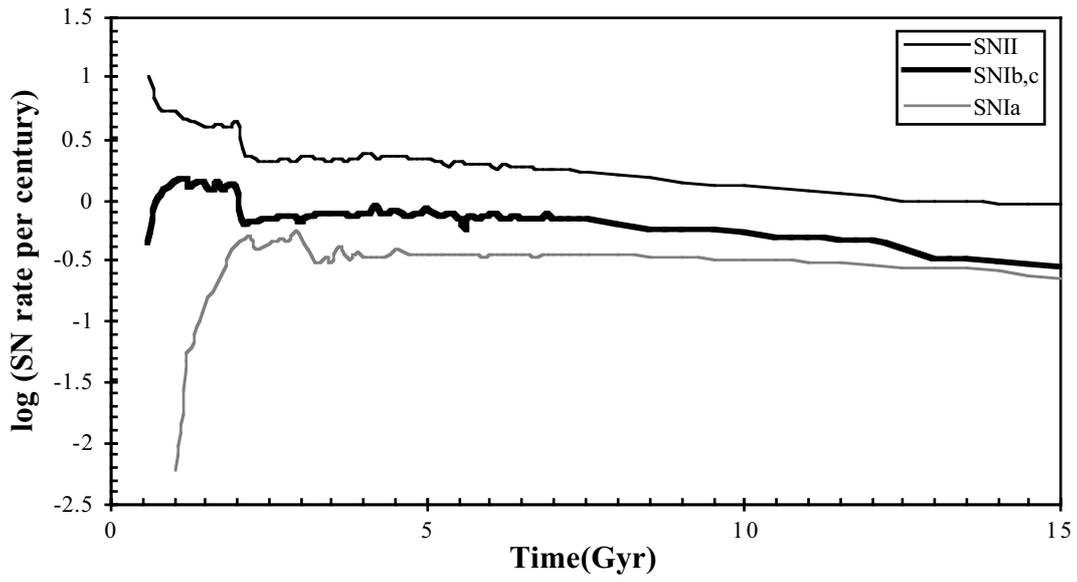

Figure 14.

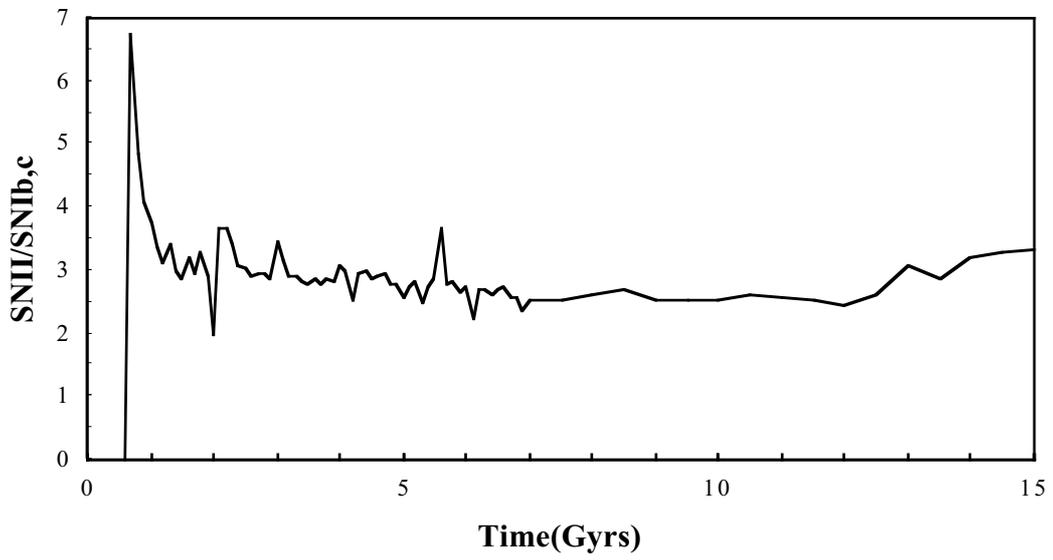

Figure 15a.

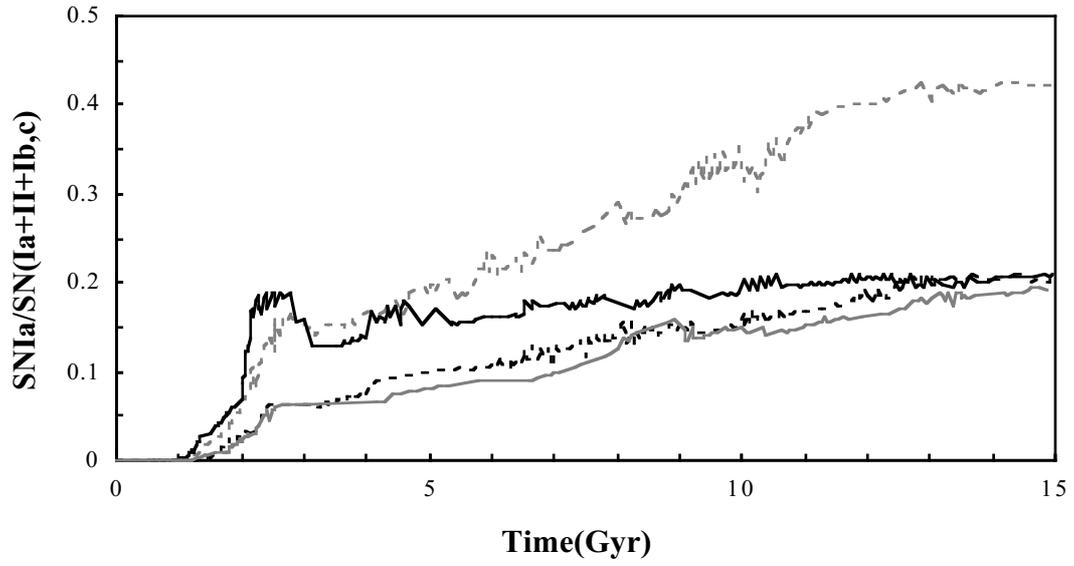

Figure 15b.